%% file: PASTIS.tex
\documentclass[usenatbib]{mn2e}

\usepackage{amsbsy}
\usepackage{graphicx}
\usepackage{txfonts}
\usepackage{epsf}
\usepackage{rotating}
\usepackage{graphics}
\usepackage{verbatim}
\usepackage{longtable}
\usepackage[hyperindex, breaklinks]{hyperref}

\def \kepler{\emph{Kepler}}

\def \MJ{M$_{\mathrm{Jup}}$}

\def \msol{M$\mathrm{_\odot}$}
\def \teff{$T_\mathrm{eff}$}
\def \kms{km\,s$^{-1}$}
\def \ms{m\,s$^{-1}$}
\def \1s{$1\,\sigma$}

\def \t0{T$_0$}

\newcommand{\prob}[2]{p(\mathrm{#1}|\mathrm{#2})}
\newcommand{\prior}[2]{\pi(\mathrm{#1}|\mathrm{#2})}

\newcommand{\aap}{A\&A}
\newcommand{\aj}{AJ}
\newcommand{\apj}{ApJ}
\newcommand{\apjl}{ApJ}
\newcommand{\apjs}{ApJS}
\newcommand{\pasp}{PASP}
\newcommand{\mnras}{MNRAS}
\newcommand{\nat}{Nature}

\begin{document}

\title[PASTIS: Bayesian exoplanet analysis and validation.]{PASTIS: Bayesian extrasolar planet validation. \\ I. General framework, models, and performance.}
        
\author[D\'iaz et al.]{R.~F.~D\'iaz,$^{1,2} $ J.~M.~Almenara,$^1$ A.~Santerne,$^{1, 3}$ C.~Moutou,$^1$ A.~Lethuillier,$^1$ M.~Deleuil$^1$\\
$^1$ Aix Marseille Universit\'e, CNRS, LAM (Laboratoire d'Astrophysique de Marseille) UMR 7326, 13388, Marseille, France\\
$^2$ Observatoire Astronomique de l'Universit\'e de Gen\`eve, 51 chemin des Maillettes, 1290 Versoix, Switzerland\\
$^3$ Centro de Astrof\'isica, Universidade do Porto, Rua das Estrelas, 4150-762 Porto, Portugal
}

\maketitle
  
\begin{abstract}

A large fraction of the smallest transiting planet candidates discovered by the \kepler\ and \emph{CoRoT} space missions cannot be confirmed by a dynamical measurement of the mass using currently available observing facilities. To establish their planetary nature, the concept of planet validation has been advanced. This technique compares the  probability of the planetary hypothesis against 
that of all reasonably conceivable alternative false-positive (FP) hypotheses. The candidate is considered as validated if the posterior probability of the planetary hypothesis is sufficiently larger than the sum of the probabilities of all FP scenarios. In this paper, we present PASTIS, the Planet Analysis and Small Transit Investigation Software, a tool designed to perform a rigorous model comparison of the hypotheses involved in the problem of planet validation, and to fully exploit the information available in the candidate light curves. PASTIS self-consistently models the transit light curves and follow-up observations. Its object-oriented structure offers a large flexibility for defining the scenarios to be compared. The performance is explored using artificial transit light curves of planets and FPs with a realistic error distribution obtained from a \kepler\ light curve. We find that data support for the correct hypothesis is strong only when the signal is high enough (transit signal-to-noise ratio above 50 for the planet case) and remains inconclusive otherwise. PLATO shall provide transits with high enough signal-to-noise ratio, but to establish the true nature of the vast majority of \kepler\ and \emph{CoRoT} transit candidates additional data or strong reliance on hypotheses priors is needed.
\end{abstract}

\begin{keywords}
planetary systems -- methods: statistical  -- techniques: photometric -- techniques: radial velocities
\end{keywords}

\section{Introduction}
Transiting extrasolar planets have provided a wealth of information about planetary interiors and atmospheres, planetary formation and orbital evolution. The most successful method to find them has proven to be the wide-field surveys carried out from the ground \citep[e.g.][]{pollacco2006, bakos2004} and from space-based observatories like \emph{CoRoT} \citep{auvergne2009} and \kepler\ \citep{koch2010}. These surveys monitor thousands of stars in search for periodic small dips in the stellar fluxes that could be produced by the passage of a planet in front of the disk of its star. The detailed strategy varies from survey to survey, but in general, since a large number of stars has to be observed to overcome the low probability of observing well-aligned planetary systems, these surveys target stars that are typically fainter than 10th magnitude.

The direct outcome of transiting planet surveys are thousands of transit light curves with depth, duration and shape compatible with a planetary transit \citep[e.g.][]{batalha2012}. However, only a fraction of these are produced by actual transiting planets. Indeed, a planetary transit light curve can be reproduced to a high level of similarity by a number of stellar systems involving binary or triple stellar systems. 
From isolated low-mass-ratio binary systems to complex hierarchical triple systems, these "false positives" are able to reproduce not only the transit light curve, but also, in some instances, even the radial velocity curve of a planetary-mass object \citep[e.g.][]{mandushev2005}.

Radial-velocity observations have been traditionally used to establish the planetary nature of the transiting object by a direct measurement of its mass\footnote{As far as the mass of the host star can be estimated, the actual mass of a transiting object can be measured without the inclination degeneracy inherent to radial-velocity measurements, since the light curve provides a measurement of the orbital inclination.}. A series of diagnostics such as the study of the bisector velocity span \citep{queloz2001}, or the comparison of the radial velocity signatures obtained using different correlation masks \citep{bouchy2009, diaz2012} are used to guarantee that the observed radial-velocity signal is not produced by an intricate blended stellar system. In addition,  these observations allow measuring the eccentricity of the planetary orbit, a key parameter for constraining formation and evolution models \citep[e.g.][]{ida2013}.

Most of the transiting extrasolar planets known to date have been confirmed by means of radial velocity measurements. However, this technique has its limitations: the radial-velocity signature of the smallest transiting companions are beyond the reach of the existing instrumentation. This is particularly true for candidates detected by \emph{CoRoT} or \kepler, whose photometric precision and practically uninterrupted observations have permitted the detection of objects of size comparable to the Earth and in longer periods than those accessible from the ground\footnote{The detection efficiency of ground-based surveys quickly falls for orbital periods longer than around 5 days \citep[e.g.][]{vonBraun2009, charbonneau2006}.}. Together with the faintness of the typical target of transiting surveys, these facts produce a delicate situation, in which transiting planets are detected, but cannot be confirmed by means of radial velocity measurements. Radial velocity measurements are nevertheless still useful in these cases to discard undiluted binary systems posing as giant planets \citep[e.g.][]{santerne2012}.

Confirmation techniques other than radial velocity measurements can sometimes be used. In multiple transiting systems, the variation in the timing of transits due to the mutual gravitational influence of the planets can be used to measure their masses \citep[for some successful examples of the application of this technique, see][]{holman2010, lissauer2011, ford2012, steffen2012, fabricky2012}. Although greatly successful, only planets in multiple systems can be confirmed this way and only mutually-resonant orbits produce large enough timing variations \citep[e.g.][]{agol2005}. Additionally, the obtained constraints on the mass of the transiting objects are usually weak.  A more generally-applicable technique is  "planet validation". The basic idea behind this technique is that a planetary candidate can be established as a \emph{bona fide} planet if the Bayesian posterior probability (i.e. after taking into account the available data) of this hypothesis is significantly higher than that of all conceivable false positive scenarios  \citep[for an exhaustive list of possible false positives see][]{santerne2013}. Planet validation is coming of age in the era of the \kepler\ space mission, which delivered thousands of small-size candidates whose confirmation by "classical" means in unfeasible. 

In this paper, we present the Planet Analysis and Small Investigation Software (PASTIS), a software package to validate transiting planet candidates rigorously and efficiently. This is the first paper of a series. We describe here the general framework of PASTIS, the modeling of planetary and false positives scenarios, and test its performances using synthetic data. Upcoming articles will present in detail the modeling and contribution of the radial velocity data (Santerne et al., in preparation), and the study of real transiting planet candidates (Almenara et al., in preparation). The rest of the article is organized as follows. In Section~\ref{sect.planetvalidation} we describe in some detail the technique of planet validation, present previous approaches to this problem and the main characteristics of PASTIS. In Section~\ref{sect.bayesian} we introduce the bayesian framework in which this work is inscribed and the method employed to estimate the Bayes factor. In Section~\ref{sect.mcmc} we present the details of the MCMC algorithm used to obtain samples from the posterior distribution. In Section~\ref{sect.priors} we briefly describe the computation of the hypotheses priors,  in Section~\ref{sect.models} we describe the models of the blended stellar systems and planetary objects.  We apply our technique to synthetic signals to test its performance and limitations in Sect.~\ref{sect.application}, we discuss the results in Section~\ref{sect.discussion}, and we finally draw our conclusions and outline future work in Section~\ref{sect.conclusions}.

\section{Planet validation and PASTIS} \label{sect.planetvalidation}

The technique of statistical planet validation permits overcoming the impossibility of confirming transiting candidates by a dynamical measurement of their mass. A transiting candidate is validated if the \emph{probability} of it being an actual transiting planet is much larger than that of being a false positive. To compute these probabilities, the likelihood of the available data given each of the competing hypothesis is needed. \citet{torres2005} constructed the first model light curves of false positives to constrain the parameters of OGLE-TR-33, a blended eclipsing binary posing as a planetary candidate that was identified as a false positive by means of the changes in the bisector of the spectral line profile. The first models of radial velocity variations and bisector span curves of blended stellar systems were introduced by \citet{santos2002}. 

In some cases, due in part to the large number of parameters as well as to their great flexibility, the false positive hypothesis cannot be rejected based on the data alone. In this situation, since the planetary hypothesis cannot be rejected either --otherwise the candidate would not be considered further--, some sort of evaluation of the relative merits of the hypotheses has to be performed, if one of them is to be declared "more probable" than the other. The concept of the probability of a hypothesis --expressed in the form of a  logical proposition-- being completely absent in the frequentist statistical approach, this comparison can only be performed through Bayesian statistics.

The BLENDER procedure \citep{torres2005, torres2011} is the main tool employed by the \kepler\ team, and it has proven very successful in validating some of the smallest \kepler\ planet candidates \citep[e.g.][]{torres2011, fressin2011, fressin2012, borucki2012, borucki2013, barclay2013}. The technique employed by BLENDER is to discard regions of the parameter space of false positives by carefully considering the \kepler\ light curve. Additional observations (either from the preparatory phase of the mission, such as stellar colors, or from follow-up campaigns, like high angular resolution imaging) are also employed to further limit the possible false positive scenarios. This is done \emph{a posteriori}, and independently of the transit light-curve fitting procedure. One of the main issues of the BLENDER tool is its high computing time \citep{fressin2011}, which limits the number of parameters of the false positives models that can be explored, as well as the number of candidates that can be studied.

\defcitealias{morton2012}{M12}
\citet{morton2012} (hereafter \citetalias{morton2012}) presented a validation procedure with improved computational efficiency with respect to BLENDER. This is accomplished by simulating populations of false positives (based on prior knowledge on Galactic populations, multiple stellar system properties, etc.), and computing the model light curve only for those "instances" of the population that are compatible with all complementary observations. Additionally, the author uses a simple non-physical model for the transit light curve (a trapezoid) independently of the model being analyzed. This is equivalent to reducing the information on the light curve to three parameters: depth, total duration, and the duration of the ingress and egress phases. Although these two features permit an efficient evaluation of the false positive probability of transiting candidates, neglecting the differences between the light curve models of competing hypotheses undermines the validation capabilities of the method in the cases where the the light curve alone clearly favours one hypothesis over the other. Although these are, for the moment, the minority of cases (Sect.~\ref{sect.koivalidation}), future space missions such as the PLAnetary Transits and Oscillations of stars (PLATO) mission will certainly change the landscape.

The approach taken in PASTIS is to obtain the Bayesian odds ratio between the planet and all false positive hypotheses, which contains all the information the data provide, as well as all the available prior information. This is the rigorous way to compare competing hypotheses. The process includes modeling the available data for the candidate system, defining the priors of the model parameters, sampling from the posterior parameter distribution, and computing the Bayesian evidence of the models. The sampling from the posterior is done using a Markov Chain Monte Carlo (MCMC) algorithm. The global likelihood or evidence of each hypothesis is computed using the importance sampling technique \citep[e.g.][]{kassraftery1995}. Once all odds ratios have been computed, the posterior distribution for the planet hypothesis can be obtained. We describe all these steps in the following sections. In Sect.~\ref{sect.blender} we perform a detailed comparison between PASTIS and the other two techniques mentioned here. 

By using a MCMC algorithm to explore the parameter space of false positives, we ensure that no computing time is spent in regions of poor fit, which makes our algorithm efficient in the same sense as the \citetalias{morton2012} method. However, the much higher complexity of the models involved in PASTIS hinders our code from being as fast as \citetalias{morton2012}. A typical PASTIS run such as the ones described in Sect.~\ref{sect.application} requires between a few hours to a few tens of hours per Markov Chain, depending on the model being used. However, these models only contain light curve data, and the modeling of follow-up observations usually requires considerable additional computing time. 

In its present state, PASTIS can model transit light curves in any arbitrary bandpass (including the \emph{CoRoT} colored light curves), the absolute photometric measurements and radial velocity data for any type of relevant false positive scenario (see Sect.~\ref{sect.models}). The models, described in Sect.~\ref{sect.models}, are as complete and as realistic as possible, to fully take advantage of the available data. A difference of our tool with respect to BLENDER and \citetalias{morton2012} is the modeling of the radial velocity data, which includes the radial velocity itself, but also the bisector velocity span of the cross-correlation function, its width and contrast. These data are very efficient in discarding false positives (see~\ref{sect.rv}). Other datasets usually available --like high angular resolution images-- are, for the moment, treated as done in BLENDER or by \citetalias{morton2012}.

%%%%%%%%%%%%
% ModelComparison  %
%%%%%%%%%%%% 
\section{Bayesian Model Comparison} \label{sect.bayesian}
The Bayesian probability theory can be understood as  an extended theory of logic, where the propositions have a degree of "plausibility" ranging between the two extremes corresponding to the classical categories "false" and "true"  \citep[see, e.g.][chapter 1]{jaynes2003}. With this definition, and unlike the frequentist approach, the Bayesian plausibility of any proposition or hypothesis\footnote{Throughout the paper, we will use the terms \emph{hypotheses} and \emph{models}. The former designate mutually-exclusive physical scenarios that can explain the observations, such as blended eclipsing binary or planetary system.  Hypotheses will be presented as logical prepositions for which Bayesian analysis is able to assign a probability. The term \emph{model} is used to designate the mathematical expressions that describe the observations. Although the two terms refer to conceptually different things, given that in our case each hypothesis will be accompanied by a precise mathematical model, we will use both terms quite freely whenever the context is sufficient to understand what is being meant.}, such as "the transit events in OGLE-TR-33 are produced by a blended stellar binary" can be precisely computed. To do this one employes the Bayes' theorem:
\begin{equation}
\prob{H_i}{D, I} = \frac{\prob{H_i}{I}\cdot\prob{D}{H_i, I}}{\prob{D}{I}}\;\;,
\end{equation}
where $\prob{X}{Y}$ is a continuous monotonic increasing function of the plausibility of preposition $X$, given that $Y$ is true \citep[see][chapter 2]{jaynes2003}. It ranges from  0 to 1, corresponding to impossibility and certainty of $\mathrm{X|Y}$, respectively. We will refer to this function as the probability of X given Y.  In principle, $H_i$, $D$, and $I$ are arbitrary propositions, but the notation was not chosen arbitrarily. Following \citet{gregory}, we designate with $H_i$ a proposition asserting that hypothesis $i$ is true, $I$ will represent the \emph{prior} information, and $D$ will designate a proposition representing the data. The probability $\prob{H_i}{I}$ is called the hypothesis prior, and $\prob{D}{H_i, I}$ is known as the evidence, or global likelihood, of hypothesis $H_i$.

The objective is to compute $\prob{H_i}{D,I}$, the posterior probability of hypothesis $H_i$, for a set of mutually-exclusive competing hypotheses $H_i$ ($i = 1, ..., N$). To proceed, it is useful to compute the \emph{odds ratio} for all pairs of hypotheses\footnote{The individual probabilities can be computed from the odds ratios, given that a complete set of hypotheses has been considered, i.e. if $\sum_{i=0}^N{\prob{H_i}{D,I}} = 1$ \citep[][chapter 3]{gregory}.}:
\begin{equation}
O_{ij} = \frac{\prob{H_i}{D, I}}{\prob{H_j}{D, I}} = \frac{\prob{H_i}{I}}{\prob{H_j}{I}}\cdot\frac{\prob{D}{H_i, I}}{\prob{D}{H_j, I}}\;\;.
\label{eq.oddsratio}
\end{equation}
The odds ratio $O_{ij}$ can therefore be expressed as a the product of two factors: the first term on the right-hand side of the above equation is known as the prior odds, and the second as the Bayes factor, $B_{ij}$. The former will be discussed in Sect.~\ref{sect.priors}, the latter is defined through:
\begin{equation}
\prob{D}{H_i, I} = \int{\prior{\boldsymbol{\theta_i}}{H_i, I}\cdot\prob{D}{\boldsymbol{\theta_i}, H_i, I}\cdot\mathrm{d}\boldsymbol{\theta_i}}\;\;,
\label{eq.evidence}
\end{equation} 
where $\boldsymbol{\theta_i}$ is the parameter vector of the model associated with hypothesis $H_i$, $\prior{\boldsymbol{\theta_i}}{H_i, I}$ is the prior distribution of the parameters, and $\prob{D}{\boldsymbol{\theta_i}, H_i, I}$ is the likelihood for a given dataset $D$.

The value of the odds ratio at which one model can be clearly preferred over the other needs to be specified. Discussions exist concerning the interpretation of the Bayes factor which are directly applicable to the odds ratio as well. Following the seminal work by \citet{jeffreys1961}, \citet{kassraftery1995} suggest interpreting the Bayes factor using a scale based on twice its natural logarithm. Thus, a Bayes factor $B_{ij}$ below 3 is considered as inconclusive support for hypothesis $i$ over hypothesis $j$, a value between 3 and 20 represents positive evidence for hypothesis $i$, between 20 and 150 the evidence is considered strong, and above 150 it is considered very strong. Because Bayesian model comparison can also provide support for the null hypothesis (i.e. model $j$ over model $i$), the inverse of these values are used to interpret the level of support for hypothesis $j$: values of $B_{ij}$ \emph{below} 1/150 indicate very strong support for hypothesis $j$.  The value of 150 has been used in the literature \citep[e.g.][]{tuomi2012}, but \citet{kassraftery1995} mention that the interpretation may depend on the context. Therefore, we will use the value of 150 as a guideline to the interpretation of the odds ratio, but we will remain flexible and will require stronger evidence if the context seems to demand it\footnote{For example, the validation of an Earth-like planet in the habitable zone of a Sun-like star naturally produces a special interest, and should therefore be treated with special care. In this case, for example, it would not be unreasonable to ask for an odds ratio above 1,000, as suggested for forensic evidence \citep[see references in][]{kassraftery1995}. In other words, "extraordinary claims require extraordinary evidence".}.

An appealing feature of the Bayesian approach to model comparison is the natural treatment of models with different numbers of parameters and of non-nested models. In this respect, Bayesian analysis has a bult-in Occam's razor that penalizes models according to the number of free parameters they have. The so-called Occam factor is included in the evidence ratio, and penalizes the models according to the fraction of the prior probability distribution that is discarded by the data \citep[see][\S 3.5, for a clear exposition on the subject]{gregory}.

\subsection{Computation of the Bayes factor} \label{sect.bayesfactor}

The evidence (equation \ref{eq.evidence}) is a $k$-dimensional integral, with $k$ equal to the number of free parameters of model $H_i$, which is in general impossible to compute analytically. We therefore approximate this integral to compute the Bayes factor using importance sampling \citep[e.g.][]{kassraftery1995}. Importance sampling is a technique to compute moments of a distribution using samples from \emph{another} distribution. Noting that equation \ref{eq.evidence} is the first moment of the likelihood function over the prior distribution $\prior{\theta}{I}$, then the evidence can be approximated as 

\begin{equation}
\prob{D}{H, I} \approx \frac{\sum_{j=1}^m w_j \prob{D}{\boldsymbol{\theta}, H, I}}{\sum_{j=1}^m w_j}\;\;,
\end{equation}
where we have dropped the hypothesis index for clarity, and the sum is done over samples of the importance sampling function $\pi^*(\theta)$, and  $w_j = \prior{\theta^{(j)}}{I}/\pi^*(\theta^{(j)} | I)$. An appropriate choice of $\pi^*(\theta)$ can lead to precise and efficient estimations of the evidence. In particular, PASTIS employs the truncated posterior-mixture estimation \citep[TPM;][]{tuomijones2012}. The importance sampling function of TPM is approximately the posterior distribution, but a small term is added to avoid the instability of the harmonic mean estimator, which uses \emph{exactly} the posterior distribution as the importance sampling function \citep[see][]{kassraftery1995}. As far as this term is small enough, samples from the posterior obtained with an MCMC algorithm (see Sect.~\ref{sect.mcmc}) can be used to estimate the evidence. \citet{tuomijones2012} show that the TPM estimator is not sensitive to the choice and size of parameter priors, a property they find convenient when comparing RV models with different number of planets, as it allows them to use very large uninformative priors (even improper priors) without penalizing their alternative models excessively. For our purposes this characteristic guarantees that the validation of a planet candidate is not a result of the choice of priors in the parameters, but rather that actual support from the data exists. As all the false positive scenarios have a larger number of degrees of freedom than the planet hypothesis, they will be severely punished by the Occam's factor. We discuss this issue further in Sect.~\ref{sect.caveats}.

\begin{figure}
\includegraphics[width=\columnwidth]{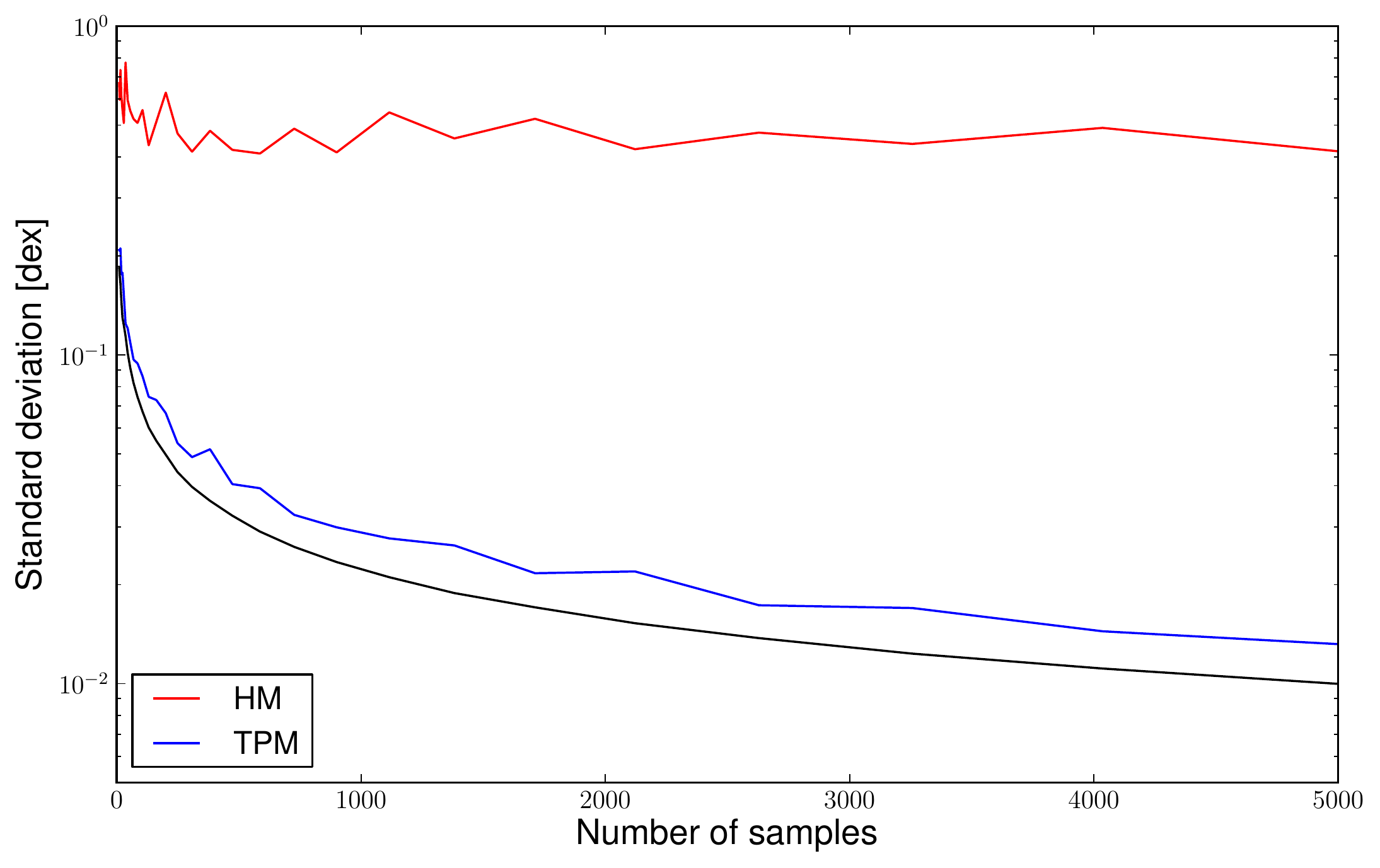}
\caption{Convergence of the truncated posterior-mixture estimator (TPM). In blue, the standard deviation of the TPM estimator as a function of sample size. The black curve is the standard deviation of the sample divided by the number of points, and the red curve is the scatter of the harmonic mean estimator (HM). It can be seen that the scatter of the TPM estimator follows approximately the black curve. For a sample size of 1000, the scatter is slightly smaller than $3\times10^{-2}$ dex. \label{fig.TPMconvergence}}
\end{figure}

The TPM estimator is supposed to verify a Gaussian central limit theorem \citep{kassraftery1995}. Therefore, as the number of independent samples ($n$) used to compute it increases, convergence to the correct value is guaranteed, and the standard deviation of the estimator must decrease as $\sqrt{n}$. In figure~\ref{fig.TPMconvergence} the standard deviation of the TPM estimator is plotted as a function of sample size, for a simple one-dimensional case for which a large number of independent samples is available. For each value of $n$, we compute the TPM and harmonic mean \citep[HM;][]{newtonraftery1994} estimators on a randomly drawn subsample. This is repeated 500 times per sample size, and in the end the standard deviation of the estimator is computed. The HM estimator (red curve) is known \emph{not} to verify a central limit theorem \citep[e.g.][]{kassraftery1995}, and indeed we see its standard deviation decreases more slowly than that of TPM. The black curve shows the mean standard deviation of the integrand of equation \ref{eq.evidence} over the selected subsample of size $n$, divided by $\sqrt{n}$. It can be seen that the TPM estimator roughly follows this curve. For our method, we require at least a thousand independent samples for each studied model, which implies a precision of around  $6\times10^{-2}$ dex in the logarithm of the Bayes factor. Given that significant support for one hypothesis over the other is given by Bayes factors of the order of 150, this precision is largely sufficient for our purposes.

Alternative methods to evaluate the Bayes factor are found in the literature. All of them are approximations to the actual computation designed to render it simpler. As such, they have their limitations. For example, the Bayesian Information Criterion (BIC), which has been widely used in the literature on extrasolar planets \citep[e.g.][]{husnoo2011}, is an asymptotic estimation of the evidence which is valid under a series of conditions. Besides requiring that all data points be independent and that the error distribution belong to the exponential family, the BIC is derived assuming the Gaussianity or near-Gaussianity of the posterior distribution \citep{liddle2007}. This means that its utility is reduced when comparing large and/or non-nested models \citep[see discussion by][]{stevenson2012}, as the ones we are dealing with here. Most of these approximations penalize models based on the number of parameters, and not on the size or shape of the prior distributions. Therefore, even parameters that are not constrained by the data are penalized. The computation of the evidence (eq. \ref{eq.evidence}), on the other hand, does not penalize these parameters, the Occam's factor being close to 1. Our aim is to develop a general method that will not depend on the data sets studied. We also want to use models with an arbitrary number of parameters, some of which will not be constrained by data but whose variations will probably contribute to the error budgets of other parameters. Furthermore, we will not be interested only in a simple ranking of hypotheses, but we will seek rather to quantify how much probable one hypothesis is over the other. Because all of this, these approximations are not useful for our purposes.

%%%%%%
% MCMC %
%%%%%%
\section{Markov Chain Monte Carlo algorithm} \label{sect.mcmc}

Markov Chain Monte Carlo (MCMC) algorithms allow sampling from an unknown probability distribution, $\prob{\boldsymbol{\theta}}{D, I}$, given that it can be computed at any point of the parameter space of interest up to a constant factor. They have been widely used to estimate the posterior distributions of model parameters \citep[see ][for a recent example]{bonfils2012}, and hence their uncertainty intervals. Here, we employ a MCMC algorithm to obtain samples of the posterior distribution that we will use to compute the evidence of different hypotheses (eq.~\ref{eq.evidence}) using the method described above. The details and many of the caveats of the application of MCMC algorithms to astrophysical problems, and to extrasolar planet research in particular, have already been presented in the literature \citep[e.g.][]{tegmark2004, ford2005, ford2006} and will not be repeated here. We do mention, on the other hand, the characteristics of our MCMC algorithm, for the sake of transparency and reproducibility.

We use a Metropolis-Hastings algorithm \citep{metropolis1953, hastings1970}, with a Gaussian transition probability for all parameters. An adaptive step size prescription closely following \citet{ford2006} was implemented, and the target acceptance rate is set to 25\%, since the problems dealt with here are usually multi-dimensional \citep[see][and references therein]{roberts1997}. %In this MCMC algorithm the probability of accepting a step from  $\boldsymbol{\theta}$ to $\boldsymbol{\theta_p}$ is given by the transition probability
%\begin{equation}
%\pi(\boldsymbol{\theta_p}, \boldsymbol{\theta}) = \min{\left(1, r\right)} = \min{\left(1, \frac{\prob{\boldsymbol{\theta_p}}{I}}{\prob{\boldsymbol{\theta}}{I}}\frac{\prob{D}{\boldsymbol{\theta_p}, I}}{\prob{D}{\boldsymbol{\theta}, I}}\right)}\;\;.
%\label{eq.acceptanceprob}
%\end{equation}
%$r$ is known as the Metropolis ratio, and the subindex $p$ indicates that $\boldsymbol{\theta_p}$ is a proposal step. The probabilities $\prob{\boldsymbol{\theta}}{I}$ and $\prob{D}{\boldsymbol{\theta}, I}$ are the prior distribution, and the likelihood for parameter vector $\boldsymbol{\theta}$, usually denoted as $\like$.

\subsection{Parameter correlations}
The parametrization of the models employed is described later, but in the most general case the parameters will present correlations that could greatly reduce the efficiency of the MCMC algorithm. To deal with this problem, we employ a Principal Component Analysis (PCA) to re-parametrize the problem in terms of uncorrelated variables. We have found that this improves significantly the mixing of our chains, rendering them more efficient. However, as already mentioned by \citet{ford2006}, only linear correlations can be fully dealt with in this way, while nonlinear ones remain a problem \footnote{We note that this is a typical problem in MCMC algorithms which have not been solved yet  and is the subject of current research \citep[e.g.][]{solonen2012}.}. To mitigate this problem, we use PCA repeatedly, i.e.\ we update the covariance matrix of the problem regularly, as described in detail at the end of this section. By doing this, the chain manages to explore "banana-shaped" correlations, as those typically existing between the inclination angle of the orbital plane and the semi-major axis, in a reasonable number of steps. This significantly reduces the correlation length \citep{tegmark2004} of the chains, producing more independent samples of the posterior for a given number of chain steps. In any case, the chains are thinned using the longest correlation length among all parameters (i.e., only one step per correlation length is kept).

\subsection{Multiple chains and non-convergence tests}
Another issue in MCMC is the existence of disjoint local maxima that can hinder the chains from exploring the posterior distribution entirely, a problem most non-linear minimization techniques share. To explore the existence of these local maxima, a number of chains (usually more than 20, depending on the dimensionality of the problem) are started at different points in parameter space, randomly drawn from the prior distributions of the parameters. Although it cannot be guaranteed that all regions of parameter space with significant posterior probability are explored, the fact that all chains converge to the same distribution is usually seen as a sign that no significant region is left unexplored. Inversely, if chains converge to different distributions, then the contribution of the identified maxima to the evidence (eq.~\ref{eq.evidence}) can be properly accounted for \citep{gregory2005}. 

To test quantitatively if our chains have converged, we employ mainly the Gelman-Rubin statistics \citep{gelmanrubin1992}, which compares the intra-chain and inter-chain variance. The chains that do not show signs of non-convergence, once thinned as explained above, are merged into a single chain that is used for parameter inference (i.e. the computation of the median value and confidence intervals), and to estimate the hypothesis evidence using the TPM method \citep{tuomijones2012}.

\subsection{Summary of the algorithm}
To summarize, our computation of the Bayes factor is done using a MCMC algorithm to sample the posterior distribution efficiently in a given \emph{a priori} parameter space. Our MCMC algorithm was compared with the \emph{emcee} code \citep{emcee} and shown to produce identical results (S.~Rodionov, priv.~comm.), with a roughly similar efficiency.

The MCMC algorithm and subsequent analysis can be summarized as follows:
\begin{enumerate}
\item Start $N_c$ chains at random points in the prior distributions. This allows exploring different regions of parameter space, and eventually find disjoint maxima of the likelihood.

\item After $N_{PCA}$ steps, the PCA analysis is started.  The covariance matrix of the parameter traces is computed for each chain, and the PCA coefficients are obtained. For all successive steps, the proposal jumps are performed in the rotated PCA space, where the parameters are uncorrelated. The value of $N_{PCA}$ is chosen so as to have around 1,000 samples of the posterior for each dimension of parameter space (i.e. to have around 1,000 different values of each parameter.

\item The covariance matrix is updated every $N_{up}$ steps, taking only the values of the trace since the last update. This allows the chain to explore the posterior distribution even in the presence of non-linear correlations.

\item The burn-in interval of each of the $N_c$ chains is computed by comparing the mean and standard deviation of the last 10\% of the chain to preceding fractions of the chain until a significant difference is found.

\item The correlation length (CL) is computed for each of the parameters, and the maximum value is retained (maxCL). The chain is thinned by keeping only one sample every maxCL. This assures that the samples in the chain are independent.

\item The Gelman \& Rubin statistics is used on the thinned chains to test their non-convergence. If the chains show no signs of not being converged, then they are merged into a single chain.

\item The TPM estimate of the evidence is computed over the samples of the merged chain.
\end{enumerate}

The whole process is repeated for all hypotheses $H_i$ of interest, such as "transiting planet" or "background eclipsing binary". The computation of the Bayes factor between any given pair of models is simply the ratio of the evidences computed in step 7.

\section{Prior odds} \label{sect.priors}
The Bayes factor is only half of the story. To obtain the odds ratio between model $i$ and $j$, $O_{ij}$, the prior odds $\prob{H_i}{D, I}/\prob{H_j}{D, I}$ in equation~\ref{eq.oddsratio} needs also to be computed. In the case of transiting planet validation, this is the ratio between the \emph{a priori} probability of the planet hypothesis and that of a given kind of false positive.

To compute the prior probability of model $H_i$ one needs to specify what is the \emph{a priori} information $I$ that is available. Note that the preposition $I$ appears as well in the parameter prior distribution, $\prior{\boldsymbol{\theta}}{I}$ in equation~\ref{eq.evidence}. Therefore, to be consistent both the parameter priors and the hypotheses priors must be specified under the same information $I$. This should be done on a case-by-case basis, but in a typical case of planet validation we usually know a few basic pieces of information about the transiting candidate: the galactic coordinates of the host star and its magnitude in at least one bandpass, and the period and depth of the transits. It is also often the case that we have information about the close environment of the target. In particular, we usually know the confusion radius about its position, i.e. the maximum distance from the target at which a star of a given magnitude can be located without being detected. This radius is usually given by the PSF of ground-based seeing-limited photometry \citep[usually the case of \emph{CoRoT} candidates; see][]{deeg2009} or by sensitivity curves obtained using adaptive optics \citep[e.g.][]{borucki2012, guenther2013}, or by the minimum distance from the star that the analysis of the centroid motion can discard \citep[mainly in the case of \kepler\ candidates; see][]{batalha2010, bryson2013}.

The  specific information about the target being studied is combined with the global prior knowledge on planet occurrence rate and statistics for different types of host star \citep[e.g.][]{mayor2011, bonfils2013, howard2010, howard2012, fressin2013}, and stellar multiple system \citep{raghavan2010}. For false positives involving chance alignments of foreground or background objects with the observed star we employ, additionally, the Besan\c con \citep{robin2003} or TRILEGAL \citep{girardi2005} galactic models to estimate the probability of such an alignment. As a test, we have verified that the Besan\c con galactic model \citep{robin2003}, combined with the three-dimensional Galactic extinction model of \citet{amoreslepine2005} reproduces the stellar counts obtained from the EXODAT catalogue \citep{deleuil2009} in a \emph{CoRoT} Galactic-center field.

\begin{table}
\begin{center}
\caption{List of model parameters.} \label{table.symbols}
\begin{tabular}{l p{6cm}}
\hline
\hline
Symbol & Parameter\\
\hline 
%\noalign{\smallskip}
\multicolumn{2}{l}{Stellar Parameters}\\
\teff & Effective temperature\\
$z$ & Stellar atmospheric metallicity \\
$g$ & Surface gravity\\
$M_{init}$ & Zero-age main sequence mass \\
$\tau_\star$ & Stellar age\\
$\rho_\star$ & Bulk stellar density \\
$v \sin i_\star$ & projected stellar rotational velocity\\
%$\mathcal{F}_\lambda$ & ... at the stellar surface\\
%$F_\lambda$ & ... outside Earth's atmosphere\\
$ua$, $ub$ & Quadratic-law limb darkening coefficients\\
$\beta$ & Gravity darkening coefficient\\
$d$ & Distance to host star\\

\noalign{\smallskip}
\multicolumn{2}{l}{Planet Parameters}\\
$M_p$ & Mass \\
$R_p$ & Radius\\
albedo & Geometric Albedo \\

\noalign{\smallskip}
\multicolumn{2}{l}{System Parameters}\\
$k_r$   & secondary-to-primary (or planet-to-star) radius ratio, $R_2/R_1$\\
%$k_r$   & secondary to primary (or planet to star) radius ratio.radius ratio between the primary (or host) star and the secondary star (or planet), $R_2/R_1$\\
$a_R$ & semi-major axis of the orbit, normalized to the radius of the primary (host) star, $a/R_1$\\
$q$ &  mass ratio, $M_2/M_1$ \\

\noalign{\smallskip}
\multicolumn{2}{l}{Orbital Parameters}\\
$P$ & orbital period\\
$T_p$ & time of passage through the periastron\\
 $T_c$ &time of inferior conjunction\\ 
$e$ & orbital the eccentricity\\
$\omega$&  argument of periastron\\
$i$ & orbital inclination  \\
$v_0$ & center-of-mass radial velocity\\
\hline
\hline
\end{tabular}
\end{center}
\end{table}

\section[sect.models]{Description of the models}\label{sect.models}
All the computations described in the previous sections require comparing the data to some theoretical model. The model is constructed by combining modeled stars and planets to produce virtually any configuration of false positives and planetary systems. The symbols used to designate the different parameters of the models are listed in Table~\ref{table.symbols}.

\begin{table*}
\begin{center}
\caption{Theoretical stellar evolutionary tracks and ranges of their basic parameters. \label{table.tracks}}
\begin{tabular}{lcccc}
\hline
\hline
Model & $M_{init}$ & step$^\dagger$& $z$ & Ref.\\
\hline
Dartmouth & [0.1, 5.0] \msol & 0.05 \msol &[-2.5, 0.5] & \citet{dotter2008}\\
Geneva & [0.5, 3.5] \msol & 0.1 \msol & [-0.5, 0.32] & \citet{mowlavi2012}\\
PARSEC & [0.1, 12] \msol & 0.05 \msol & [-2.2, 0.7] & \citet{bressan2012}\\
StarEvol & [0.6, 2.1] \msol & 0.1 \msol &[-0.5, 0.5] & Palacios (priv. comm.)\\
\hline
\hline
\end{tabular}
\end{center}
\small{$\dagger$} Grid step is not constant thorough grid range. Typical size is reported.
\end{table*}

\begin{table*}
\begin{center}
\caption{Theoretical stellar spectra. \label{table.spectra}}
\begin{tabular}{lcccc}
\hline
\hline
Model & $z$ & \teff & $\log g$ & Ref.\\
\hline
ATLAS/Castelli \& Kurucz & [-2.5, 0.5] & [3500, 50000] K & [0.0, 5.0] cgs & \citet{castellikurucz2004}\\
PHOENIX/BT-Settl & [-4.0, 0.5] & [400, 70000] K & [-0.5, 6.0] cgs & \citet{allard2012} \\
\hline
\hline
\end{tabular}
\end{center}
\end{table*}

\subsection{Modeling stellar and planetary objects}
Planetary objects are modeled as non-emitting bodies of a given mass and radius, and with a given geometric albedo. To model stellar objects we use theoretical stellar evolutionary tracks to obtain the relation between the stellar mass, radius, effective temperature, luminosity, metallicity, and age. The theoretical tracks implemented in PASTIS are listed in Table~\ref{table.tracks}, together with their basic properties. Depending on the prior knowledge on the modeled star, the input parameter set can be either [\teff, $\log \mathrm{g}$, $z$], [\teff, $\rho$, $z$], or [$M_{init}$, Age, $z$]. In any case, the remaining parameters are obtained by trilinear interpolation of the evolution tracks. 

Given the stellar atmospheric parameters \teff, $\log \mathrm{g}$, and $z$, the output spectrum of the star is obtained by linear interpolation of synthetic stellar spectra (see Table~\ref{table.spectra}). The spectrum is scaled to the correct distance and corrected from interstellar extinction:
\begin{equation}
F_\lambda = \mathcal{F}_\lambda \, d^{-2} \, 10^{-0.4\cdot R_\lambda\cdot E(B-V)}\;\;,
\end{equation}
where $\mathcal{F}_\lambda$ is the flux at the stellar surface of the star and $F_\lambda$ is the flux outside Earth's atmosphere. $R_\lambda = A_\lambda/E(B-V)$ is the extinction law from \citet{fitzpatrick1999} with $R_V = 3.1$, and $E(B-V)$ is the color excess, which depends on the distance $d$ to the star and on its galactic coordinates. In PASTIS, $E(B-V)$ can be either fitted with the rest of the parameters or modeled using the three-dimensional extinction model of \citet{amoreslepine2005}. The choice to implement different sets of stellar tracks and atmospheric models allows us to study how our results change depending on the employed set of models, and therefore to estimate the systematic errors of our method.

The spectra of all the stellar objects modeled for a given hypothesis are integrated in the bandpasses of interest to obtain their relative flux contributions \citep{bayo2008}. They are then added together to obtain the total observed spectrum outside Earth's atmosphere, from which the model of the observed magnitudes is likewise computed. Note that by going through the synthetic spectral models rather than using the tabulated magnitudes from the stellar tracks (as is done in BLENDER), any arbitrary photometric bandpass can be used, as long as its transmission curve and its flux at zero magnitude are provided. In particular, this allows us to consider the different color \emph{CoRoT} light curves \citep{rouan2000, rouan1998} that should prove a powerful tool to constrain false positive scenarios (Moutou et al. submitted).

Additional parameters of the star model are the limb-darkening coefficients, and the gravity-darkening coefficient $\beta$, defined so that \teff$^4 \propto g^\beta$.
The limb-darkening coefficients are obtained from the tables by \citet{claret2011} by interpolation of the atmospheric parameters \teff, $\log \mathrm{g}$, and $z$. Following \citet{espinosalara2012}, the gravity-darkening coefficient is fixed to 1.0 for all stars. Of course, these coefficients can also be included as free parameters of the model at the cost of potential inconsistencies, such as limb-darkening coefficients values that are incompatible with the remaining stellar parameters.

\subsection{Modeling the light curve and radial velocity data}
The light curves of planets and false positives are modeled using a modified version of EBOP code \citep{nelsondavis1972, etzel1981, popperetzel1981} which was extracted from the JKTEBOP package \citep[][and references therein]{southworth2011}. The model parameters can be divided in achromatic parameters, which do not depend on the bandpass of the light curve, and chromatic ones.  The set of achromatic parameters chosen are: [$k_r$, $a_R$, $i$, $P$, $T_p$, $e$, $\omega$, $q$]. In some cases, we use the time of inferior conjunction $T_c$ instead of $T_p$, because it is usually much better constrained by observations in eclipsing (transiting) systems. The mass ratio $q$ is among these parameters because EBOP models ellipsoidal modulation, which allows us to use the full-orbit light curve to constrain the false positive models. Additionally, we included the Doopler boosting effect \citep[e.g.][]{faigler2012} in EBOP. The chromatic parameters are the coefficients of a quadratic limb-darkening law ($ua$ and $ub$), the geometric albedo, the surface brightness ratio (in the case of planetary systems, this is fixed to 0), and the contamination factor due to the flux contribution of nearby stars inside the photometric mask employed.

The model light curves are binned to the sampling rate of the data when this is expected to produce an effect on the adjusted parameters \citep{kipping2010}. For blended stellar systems, the light curves of all stars are obtained, they are normalized using the fluxes computed from the synthetic spectra as described above, and added together to obtain the final light curve of the blend.

The model for radial velocity data is fully described in Santerne et al. (in preparation). Briefly, the model constructs synthetic cross-correlation functions (CCFs) for each modeled star using the instrument-dependent empirical relations between stellar parameters ($z$ and  $B-V$ color index) and CCF contrast and width obtained by \citet{santos2002}, \citet{boisse2010}, and additional relations described in Santerne et al.~(in prep.). Our model assumes that each stellar component of the modeled system contributes to the observed CCF with a Gaussian profile located at the corresponding radial velocity, and scaled using the relative flux of the corresponding star. The resulting CCF is fitted with a Gaussian function to obtain the observed position, contrast and full-width at half-maximum. The CCF bisector is also computed \citep{queloz2001}. As planetary objects are modeled as non-emitting bodies, their CCF is not considered.

\subsection{Modeling of systematic effects in the data} \label{sect.jitter}
In addition, we use a simple model of any potential systematic errors in the data not accounted for in the formal uncertainties. We follow in the steps of \citet{gregory2005}, and model the additional noise of our data as a Gaussian-distributed variable with variance $s^2$. The distribution of the total error in our data is then the convolution of the distribution of the known errors with a Gaussian curve of width $s$. When the known error of the measurements can be assumed to have a Gaussian distribution\footnote{The method being described is not limited to treat gaussian-distributed error bars. In fact, any arbitrary distribution can be used without altering the algorithm and models described so far. Only the computation of the likelihood has to be modified accordingly.} of width $\sigma_i$, then the distribution of the total error is also a Gaussian with a variance equal to $\sigma_i^2 + s^2$.

In principle, the additional parameter $s$ is uninteresting and will be marginalized. \citet{gregory} claims that this is a robust way to obtain conservative estimates of the posterior of the parameters. Indeed, we have found that in general, adding this additional noise term in the MCMC algorithm produces wider posterior distributions. 

\subsection{The false positive scenarios} \label{sect.fp}

The modeled stellar and planetary objects can be combined arbitrarily to produce virtually any false positive scenario. For single-transit candidates, the relevant models that are constructed are:
\begin{itemize}
\item Diluted eclipsing binary. The combination of a target star, for which prior information on its  parameters [\teff, $\log \mathrm{g}$, $z$] is usually available, and a couple of blended stars in an eclipsing binary (EB) system with the same period as the studied candidate. Usually, no \emph{a priori} information exists on the blended stars because they are much fainter than the target star. Therefore they are parametrized using their initial masses $M_{init}$, and the age and metallicity of the system (the stars are assumed to be co-eval). The diluted EB can be located either in the foreground or in the background of the target star.
\smallskip
\item Hierarchical triple system. Similar to the previous case, but the EB is gravitationally bound to the target star. As a consequence, all stars share the same age and metallicity, obtained from the prior information on the target star.
\smallskip
\item Diluted transiting planet. Similar to the diluted eclipsing binary scenario, but the secondary star of the EB is replaced by a planetary object.
\smallskip
\item Planet in Binary. Similar to the hierarchical triple system scenario, but the secondary star of the EB is replaced by a planetary object.
\end{itemize}
In addition, the models involving a diluted eclipsing binary should also be constructed using a period twice the nominal transit period, but these scenarios are generally easily discarded by the data \citep[e.g.][]{torres2011}. Undiluted eclipsing binaries may also constitute false positives, in particular those exhibiting only a secondary eclipse \citep{santerne2013}, and are also naturally modeled by PASTIS. However, since they can be promptly discarded by means radial velocity measurements, they are not listed here and are not considered in Sect.~\ref{sect.application}.
Finally, the transiting planet scenario consists of a target star orbited by a planetary object. In this case, it is generally more practical to parametrize the target star using the parameter set [\teff, $\rho_\star$, $z$], where the stellar density $\rho_\star$ replaces the surface gravity $\log g$, since it can be constrained from the transit curve much better.

For candidates exhibiting multiple transits, the number of possible models is multiplied because any given set of transits can in principle be of planetary or stellar origin. PASTIS offers a great flexibility to model false positives scenarios by simply assembling the basic "building blocks" constituted by stars and planets.

\section{Application to synthetic light curves}\label{sect.application}

\begin{table}
\begin{center}
\caption{Parameters for synthetic light curves \label{table.syntheticparams}}
\begin{tabular}{l c}
\hline
\hline
\multicolumn{2}{l}{Transiting Planet}\\
\hline
%\noalign{\smallskip}
%Planet Radius [R$_\odot$]& $\{0.0092; 0.0404; 0.0716; 0.1028\}$\\
Planet Radius [R$_\oplus$]& $\{1.0; 4.4; 7.8; 11.2\}$\\
Impact Parameter $b$ & $\{0.0; 0.5; 0.75\}$\\
Transit S/N & $\{10; 20; 50; 100; 150; 500\}$\\
\hline
\hline
%\noalign{\smallskip}
\multicolumn{2}{l}{Background Eclipsing Binary}\\
\hline
%\noalign{\smallskip}
Mass Ratio& $\{0.1; 0.3; 0.5\}$\\
Impact Parameter $b$ & $\{0.0; 0.5; 0.75\}$\\
Secondary S/N & $\{2; 5; 7\}$\\
\hline
\end{tabular}
\end{center}
\end{table}

This section explores the capabilities and limitations of our method. We inject synthetic signals of planets and background eclipsing binaries (BEBs) in real \kepler\ data, and use it to run the validation procedure. In each case, both the correct and incorrect model are tested, and the odds ratio for these two scenarios is computed. We will refer to the models used to fit the data as the PLANET and BEB models. For the sake of simplicity, we did not include radial velocity or absolute photometric data, although they are important in the planet validation process of real cases \citep[][and Sect.~\ref{sect.rv}]{ollivier2012}. Only light curve data is modeled in this Section. We describe the synthetic data and models in Sect.~\ref{sect.synthLC}. In Sections \ref{sect.PLAresults} and \ref{sect.BEBresults}, we study what type of support is given to the correct hypothesis by the data, independently of the hypotheses prior odds. To do this, we compute the Bayes factor -- i.e. the second term in right-hand side of Eq.~\ref{eq.oddsratio}. Finally, in Sect.~\ref{sect.inclHypPriors} we compute the odds ratio by assuming the target environment and follow-up observations of a typical \kepler\ candidate. In Sect.~\ref{sect.otherFP} we study the remaining false positive scenarios described in Sect.~\ref{sect.fp} 

\subsection{Synthetic light curves and models \label{sect.synthLC}}

\begin{figure}
\includegraphics[width = \columnwidth]{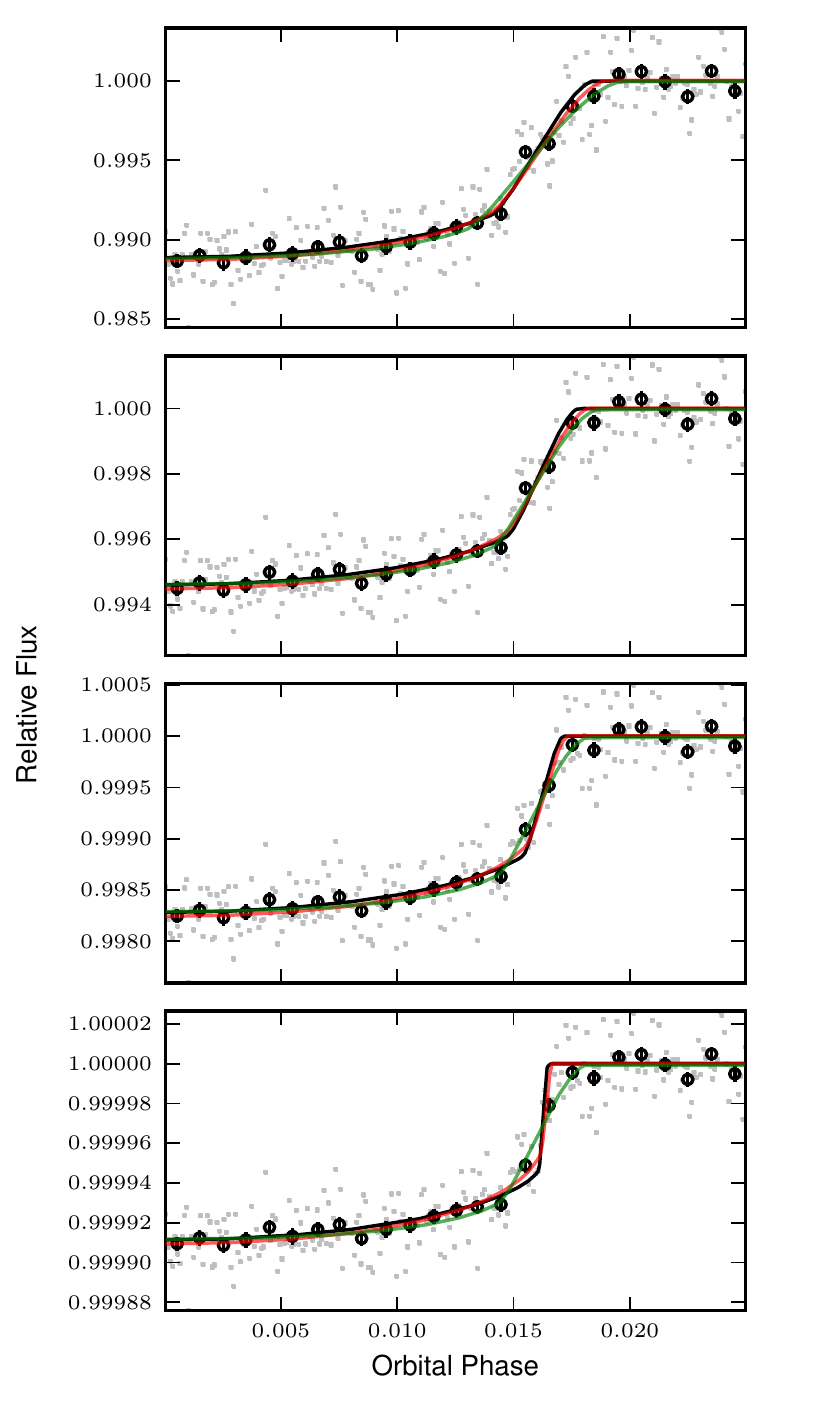}
\caption{Synthetic planetary egress transit light curves with S/N = 150 and $b = 0.5$ for the four sizes of simulated planets (see Table~\ref{table.syntheticparams}).The grey dots are the individual binned data points (see text), the black circles represent the average of the data in 0.001-size bins. The red and green curves are the maximum-posterior PLANET and BEB models, respectively, found using the MCMC algorithm. The black curve is the actual model injected in the \kepler\ light curve (barely distinguishable from the PLANET best model). These figures show the effect of the size of the planet --the change of ingress/egress duration-- independently of the S/N of the light curve. Note the change of the y-axis scale as we go from a Jupiter-size planet to an Earth-size planet. Note also the excess of points below the injected model around phase 0.019.}
\label{fig.transits}
\end{figure}

Photometric data from the \kepler\ mission have the best precision available to date. More relevant for our tests, the instrument is extremely stable. Indeed, by design, instrumental effects do not vary significantly in the timescale of planetary transits \citep{koch2010}. As a consequence, \kepler\ data is well suited for planet validation, because systematic effects are not expected to reproduce small transit features that could unfairly favor false-positive scenarios. By choosing to use a \kepler\ light curve as a model for the error distribution of our synthetic data we are considering the best-case scenario for planet validation. However, we shall see in Sect.~\ref{sect.PLAresults} that a small systematic effect in the light curve has a significant effect on the results.

To test our method in different conditions of signal-to-noise (S/N), transit shape, and dilution, the light curves of transiting planets and BEBs were generated with different parameter sets using the models described above. The parameters sets are presented in Table~\ref{table.syntheticparams}. In all cases, the period of the signal is 3 days, and the orbit is circular. The synthetic signals were injected in the \kepler\ short-cadence data of star KIC11391018. This target has a magnitude of 14.3 in the \kepler\ passband, which is typical for the transiting candidates that can be followed-up spectroscopically from the ground \citep[e.g.][]{santerne2012}. Its noise level, measured with the rms of the Combined Differential Photometric Precision (CDPP) statistics over 12 hours, is near the median of the distribution for stars in the same magnitude bin (i.e.\ $Kp$ between 13.8 and 14.8) observed in Short Cadence mode in Quarter 4. These two conditions make it a typical star in the \kepler\ target list. On the other hand, it is located in the 82nd percentile of the noise level distribution of \emph{all} Long Candence \kepler\ target stars in this magnitude bin, demonstrating a bias towards active stars in the Short Cadence target list. Additionally, KIC11391018 exhibits planetary-like transits every around 30 days (KOI-189.01), which were taken out before injecting the model light curves. To reduce the computation time spent in each parameter set, the synthetic light curves were binned to 10,000 points in orbital phase. This produces an adequate sampling of transit, and enough points in the out-of-transit part. Note that since the transit signals are injected in the short-cadence data the sampling effects described in, for example, \citet{kipping2010} are not present here. If needed PASTIS deals with this issue oversampling the light curve model and then binning back to the observed cadence before comparing with the data, as done, for example, in \citet{diaz2013} and \citet{hebrard2013}. Short-cadence data was chosen because it resembles the cadence of the future PLATO data \citep{rauer2013}, but this should be taken into account when interpreting the results from the following sections (see also Section~\ref{sect.caveats}).

For the target star (i.e. the planet host star in the PLANET model, and the foreground diluting star in the BEB model), we chose a 1-M$_\odot$, 1-R$_\odot$ star. We assumed that spectroscopic observations of the system have provided  information on atmospheric parameters of this star ($T_\mathrm{eff}$, $z$, and $\log g$). For the BEB model, we assume that no additional information is available about the background binary. These stars are therefore modeled using their initial mass, metallicity and age.
To consistently model the radii, fluxes, and limb darkening coefficients of the stars involved in the models, we used the Dartmouth evolution tracks and the Claret tables as explained above.

For each synthetic light curve, the PASTIS MCMC algorithm was employed to sample from the parameter joint posterior. For simplicity, the orbital period, and epoch of the transits / eclipses were fixed to the correct values; because light curve data is incapable of strongly constraining the eccentricity, we fixed it to zero; we chose a linear limb darkening law. 

The model parameters and the priors are listed in Table \ref{table.paramspriors}. Note that the age, metallicity, and albedo of the background binary star are also free parameters of our model, even if we do not expect to constrain them. For the target star, we chose priors on $T_\mathrm{eff}$, $z$, and $\log g$ as those that would be obtained after a spectroscopic analysis. In the PLANET model, the bulk stellar density $\rho_*$ obtained from the Dartmouth tracks is used instead of $\log g$.  For the masses of the background stars we used the initial mass function (IMF) used in the Bensa\c con Galactic model \citep{robin2003} for disk stars. We also assumed the stars are uniformly distributed in space, hence a $d^2$ prior for the distance. Uninformative priors were used for all remaining parameters.  The current knowledge on the radius distribution of planets is plagued with biases, and suffers from the incompleteness of the transiting surveys, and other systematic effects. We therefore preferred a log-flat prior to any possible informative prior. Additionally, this functional form is not far from the planet radius distribution emerging from the \kepler\ survey \citep{howard2012, fressin2013}. Lastly, for the BEB model we required the \emph{ad hoc} condition that the binary be at least one magnitude fainter than the target star. We assumed that if this condition were not fulfilled the binary would be detectable, and the system would not be considered a valid planetary candidate.

Ten independent chains  of 700,000 steps were run for each synthetic dataset, starting at random points drawn from the joint prior distribution. After trimming the burn-in period and thinning the chains using their correlation length, we required a minimum of 1000 independent samples (see Sect.~\ref{sect.bayesfactor}). When this was not fulfilled, additional MCMCs were run to reach the required number of samples. The number of independent samples obtained for each simulation is presented in Tables~\ref{table.resultsPLA} and \ref{table.resultsBEB}. The evidence of each model was estimated using the MCMC samples as explained in Sect.~\ref{sect.bayesfactor}. In the next two sections we present the results of the computation of the Bayes Factor for the planet and BEB synthetic light curves.

\begin{table}
\caption{Jump parameters and priors of the planet and BEB models used to fit the synthetic light curves. \label{table.paramspriors}}
\begin{center}
\begin{tabular}{ll}
\hline
\hline
%\noalign{\smallskip}
\multicolumn{2}{l}{Common parameters}\\
\hline
Target $T_\mathrm{eff}$ [K]	& Normal(5770, 100)	\\
Target metallicity, $z$ [dex]	& Normal(0.0, 0.1)	\\
Systematic noise, $\sigma_J$ [ppm]		& Uniform(0, 3$\sigma$)$^\dagger$\\
Out-of-transit flux			& Uniform(1 - $\sigma$, 1 + $\sigma$)$^\dagger$\\
\hline
\hline
%\noalign{\smallskip}
\multicolumn{2}{l}{\emph{PLANET model}}\\
\hline
%\noalign{\smallskip}
Target density $\delta_*$	[solar units]			& Normal(0.93, 0.25)\\
$k_r = R_p/R_1$				& Jeffreys(10$^{-3}$, 0.5)\\
Planet albedo					& Uniform(0.0, 1.0) \\
Orbital inclination $i$  [deg]		& Sine(80, 90) \\
%\noalign{\smallskip}
\hline\hline
%\noalign{\smallskip}
\multicolumn{2}{l}{\emph{BEB model}}\\
\hline
%\noalign{\smallskip}
Target $\log g$	[cgs]			& Normal(4.44, 0.1)\\
Primary / Secondary $M_{init}$ [\msol]	& IMF$^*$	\\
Primary / Secondary albedo	& Uniform(0.6, 1.0)\\
Binary $\log \tau_\star$ [Gyr]	& Uniform(8, 10)	\\
Binary $z$ [dex]			& Uniform(-2.5, 0.5) 	\\
Binary distance	, $d$ [pc]		& $d^2$, for $d \in [0 - 5000]$	\\
Impact parameter $b$		& Uniform(0.0, 1.0)		\\
%\noalign{\smallskip}

\hline
\end{tabular}
\end{center}
\small{$^*$:} the initial mass function was modeled as the disk IMF in the Bensan\c con Galactic model \citep{robin2003} with a segmented power law $\mathrm{d} n/ \mathrm{d} m \propto m^{-\alpha}$, with $\alpha = 1.6$ if $m < 1.0$ \msol, and $\alpha = 3.0$ if $m > 1.0$ \msol.

\small{$\dagger$:} $\sigma$ represents the mean uncertainty of the light curve data.
\end{table}

\begin{figure*}
\begin{center}
\includegraphics[height = 0.38\textwidth]{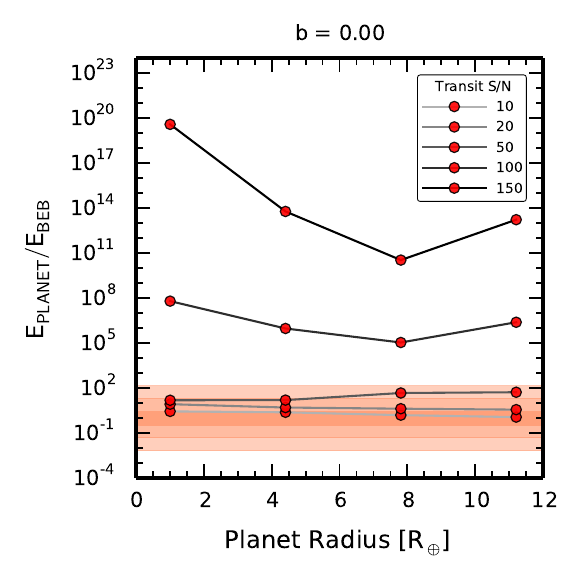}
\includegraphics[height = 0.38\textwidth]{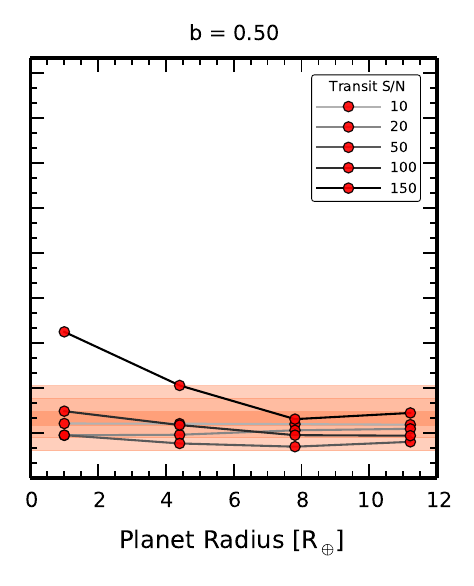}
\includegraphics[height = 0.38\textwidth]{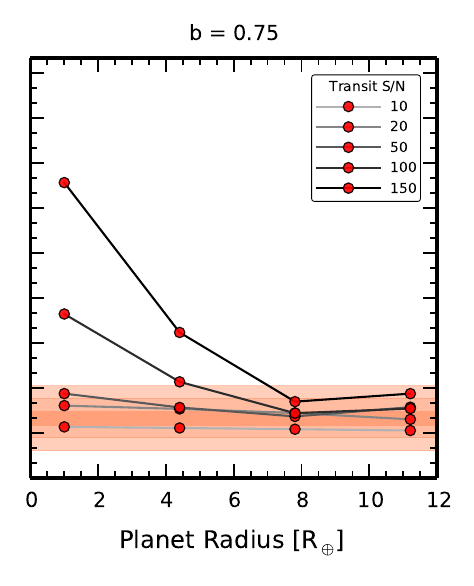}
\end{center}
\caption{Evidence ratio (Bayes factor) between PLANET and the BEB model for the synthetic planet data with impact parameter b = 0.0 (left), b = 0.5 (center), and b = 0.75 (right). In each panel, the results are plotted in logarithmic scale as a function of the simulated planetary radius for different signal-to-noise ratios of the transit: from 10 (light gray) to 150 (black). The 68.3\% confidence interval is smaller than the size of the symbols. The shaded areas indicate the regions where the support of one model over the other is arbitrarily considered "Inconclusive" (dark), "Positive", "Strong" (light orange), and "Very strong" (white), according to \citet{kassraftery1995}.
}

\label{fig.resultsPLANET}
\end{figure*}

\subsection{Planet simulations \label{sect.PLAresults}}
The light curves of planetary transits were constructed for different values of the radius of the planet, the impact parameter ($b$) and the S/N of the transit, defined as
\begin{equation}  
S/N = \frac{\delta_0}{\sigma}\sqrt{N_t}\;\;,
\label{eq.snr}
\end{equation}
where $\delta_0$ is the fractional depth of the central transit, $\sigma$ is the data scatter (measured outside the transit), and $N_t$ is the number of points inside the transit. The values of the parameters are shown in Table~\ref{table.syntheticparams}. We explore planets with sizes ranging between the Earth's and Jupiter's, transits with impact parameters between 0.0 and 0.75, and S/N ranging between 10 and 150 \footnote{The \kepler\ candidates with estimated radius below 1.4 R$_\oplus$ have mean S/N = 20. } In total, 72 different transiting planet light curves were analyzed.

For a given star, reducing the size of the planet changes both the shape of the transit and its S/N. To correctly disentangle the effects of the size of the planet and of the S/N of the transit, light curves with different S/N were constructed for a given planet radius. Although transits of Earth-size planets rarely have S/N of 150 among the \kepler\ candidates (see Sect.~\ref{sect.koivalidation}), these type of light curves should be more common in the datasets of the proposed space mission PLATO\footnote{The S/N of a transit of an Earth-size planet in front of a 11th-magnitude 1-R$_\odot$ star over 2 years of continuous observations with PLATO should be around 450 and 60 for periods of 10 and 100 days, respectively. PLATO will observe around 20,000 stars brighter than 11th magnitude for at least this period of time. The observing campaigns of the future Transiting Exoplanet Survey Satellite (TESS) being shorter, the obtained S/N will we lower, except for very few stars near the celestial poles.}, because it will target much brighter stars for equally-long periods of time. To modify the S/N of the transits, the original light curve was multiplied by an adequate factor. We estimate this factor for the central transits (i.e. with $b = 0$), and used it for constructing the light curves with $b = 0.50$ and $b =0.75$. This produces a somewhat lower S/N for these transits, both due to the fewer number of points during the transit and the shallower transit. The S/N of transits with $b = 0.75$ is reduced by about 20\% with respect to those with $b  = 0.0$. A few examples of the synthetic planetary transit light curves are shown in Figure~\ref{fig.transits}. 

\begin{figure}
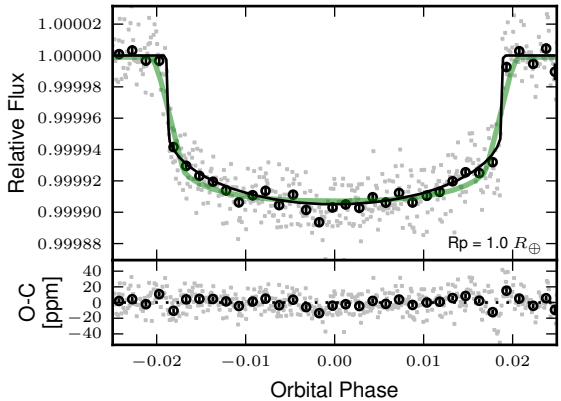

\center
\includegraphics[width = \columnwidth]{{{plots/PLANET_transit_residuals_1Rearth_b0.00_snr150}}}
\caption{Example of the impossibility of the BEB model to reproduce sharp ingresses or egresses. The transit light curve of the simulated Earth-size planet with S/N = 150, and $b = 0.0$ is shown together with the maximum-posterior BEB model (green curve) and the injected model (black curve). The black points represent the average of the data points over 0.15\% of the orbital phase. The residuals to the best fit are shown in the lower panel. The error bars include the maximum-posterior systematic noise amplitude (5.3 ppm; see Sect.~\ref{sect.jitter})  \label{fig.residuals}}
\end{figure}

The results of the evidence computation are summarized in Table~\ref{table.resultsPLA}, where columns 1 to 3 list the values of the parameters used to construct the synthetic light curve, columns 4 and 5 are the number of independent samples obtained for each model after trimming the burn-in period and thinning using the correlation length, column 6 is the logarithm (base 10) of the Bayes factor, column 7 is the logarithm of the final odds ratio; columns 8 and 9 are the 64.5\% upper and lower confidence levels for the Bayes Factor. The following columns are used to perform frequentist tests described in Sect.~\ref{sect.frequentist}: columns 10 and 11 are the reduced $\chi^2$ for each model, column 12 is the F-test \emph{p}-value, column 13 is the likelihood-ratio test statistics, and column 14 is the \emph{p}-value of this test (computed only if $D > 0.0$).  The results are plotted in Figure~\ref{fig.resultsPLANET}, where the Bayes factor in favour of the PLANET model, $B_\mathrm{PB} = \prob{D}{H_\mathrm{PLANET}, I}/\prob{D}{H_\mathrm{BEB}, I} = E_\mathrm{PLANET}/ E_\mathrm{BEB}$, is plotted as a function of the radius of the simulated planet and the transit S/N. The uncertainties were estimated by randomly sampling (with replacement) $N$ elements from the posterior sample obtained with MCMC, where $N$ is the number of independent steps in the chain. The Bayes factor was computed on 5,000 synthetic samples generated thus, from which the 68.3\% confidence regions are obtained. In the plots, the uncertainties are always smaller than the size of the symbols.
The shaded regions show the limiting values of $B_\mathrm{PB}$ described in Sect.~\ref{sect.bayesian}. The lightest shaded areas extend from 20 to 150, above which the support for the PLANET model is considered as "very strong", and between 1/150 and 1/20, below which the support is considered "very strong" for the BEB model.

It can be seen that the Bayes factor increases rapidly with the S/N. {For the highest S/N},  $B_\mathrm{PB}$ decreases with $R_p$ from 1 to 8 $R_\oplus$, and increases again slightly for Jupiter-size planets, for which the \emph{ad-hoc} brightness condition becomes relevant. For the low S/N simulations, the dependence with the planet radius is less clear but roughly follows the same trend. Because the duration of the ingress and the egress becomes shorter as the size of the planet decreases, the BEB model is unable to correctly reproduce the light curve of Earth- and Neptune-size planets for the simulations with S/N $> 50$ (Fig.~\ref{fig.residuals}), but both models are statistically undistinguishable for lower S/N transits. \footnote{See also the example of the Q1-Q3 transit of Kepler-9 d in \citet[][Fig. 11]{torres2011}; by Quarter 6 the transit had a S/N $\sim 13$ only (\url{http://nexsci.caltech.edu/}).}. In any case, all fitted models are virtually equally "good". This can be seen in Table~\ref{table.resultsPLA}, where we list the reduced $\chi^2$ of the best-fit model of each scenario, computed including the systematic error contribution obtained with PASTIS. The fact that all values are close to unity imply that a frequentist test will fail to reject \emph{any} of the models explored here. We discuss this in more detail in Sect.~\ref{sect.frequentist}. Note that in no case the Bayes factor gives conclusive support for the BEB model, even though it somehow favours it for $b = 0.5$.

In this regard, a monotonic decrease of $B_\mathrm{PB}$ with impact parameter $b$ was expected. However, the Bayes factor decreases from $b = 0.0$ to $b = 0.5$, and it increases again as the transit becomes less central. Additionally, the synthetic light curves with $b = 0.5$ are fitted better (i.e., the likelihood distribution is significantly shifted towards larger values) than the corresponding light curves with $b = 0.0$ and $b = 0.75$ both for the PLANET and BEB models. There should be no reason why the PLANET model fits  the light curves with $b = 0.5$ better. The cause of the observed decrease in $B_\mathrm{PB}$ has to be a feature of the light curve not produced by the synthetic models. An inspection of the light curves, the maximum-posterior curves for each model, and the evolution of the merit function across the transit reveals that this is due to a systematic distorsion of the light curve occurring at phase $\sim 0.019$ (see Fig.~\ref{fig.transits}). This feature produce a two-folded effect that explains the decrease of the Bayes factor for $b = 0.5$. Firstly, at $b = 0.0$ the "bump" occurs near the egress phase, which increases the inadequacy of the BEB model to reproduce the transit of small planets (see Fig.~\ref{fig.residuals}, where the residuals are asymmetric between ingress and egress). This increases the Bayes factor for the PLANET model for $b = 0.0$, specially for small planets. Secondly, at $b = 0.5$ the distorted phase is just outside or at fourth contact for small and giant planets, respectively. The BEB model produces a better fit because the egress duration is larger (see Fig.~\ref{fig.transits}). As mentioned above, the PLANET model fits the data better as well, but the improvement is less dramatic; note that the maximum-posterior transit duration is systematically larger than that of the injected transits (Fig.~\ref{fig.transits}). As a consequence, the Bayes factor is reduced significantly for $b = 0.5$.  Finally, as the systematic "bump" is well outside the transit for $b = 0.75$ it does not produce an artificial increase of the likelihood of any of the models. We believe this systematic effect explains the unexpected dependence of $B_\mathrm{PB}$ with impact parameter.

A corollary of this discussion is that the simple modeling of systematics effects as and additional source of Gaussian noise is not sufficient to treat \kepler\ data. Under a correct noise model, the current maximum-posterior model should not be preferred over the actual injected model. This clearly signals a line of future development in PASTIS.

Light curves with S/N = 500 were likewise studied, but their results do not appear in the figures nor in the tables above. The reason for this is that they produce overwhelming support for the correct hypothesis, for all planetary radii and impact parameters. Providing the exact value of the Bayes factor for these cases was not deemed useful.

\begin{figure*}
\includegraphics[width = \textwidth]{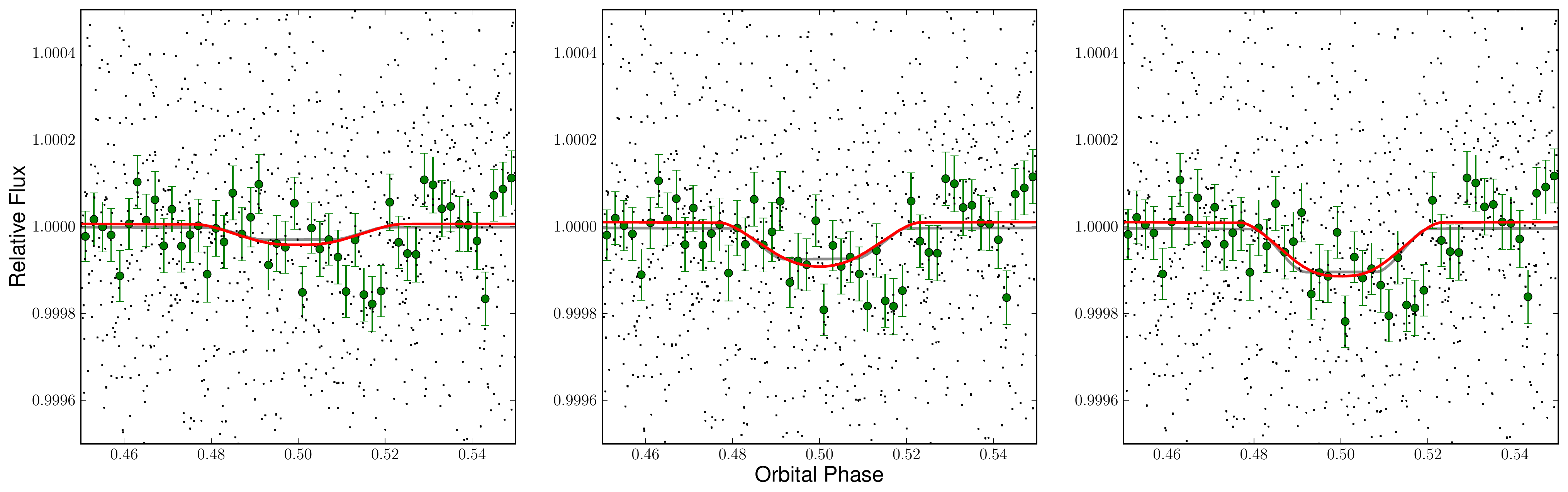}
\includegraphics[width = \textwidth]{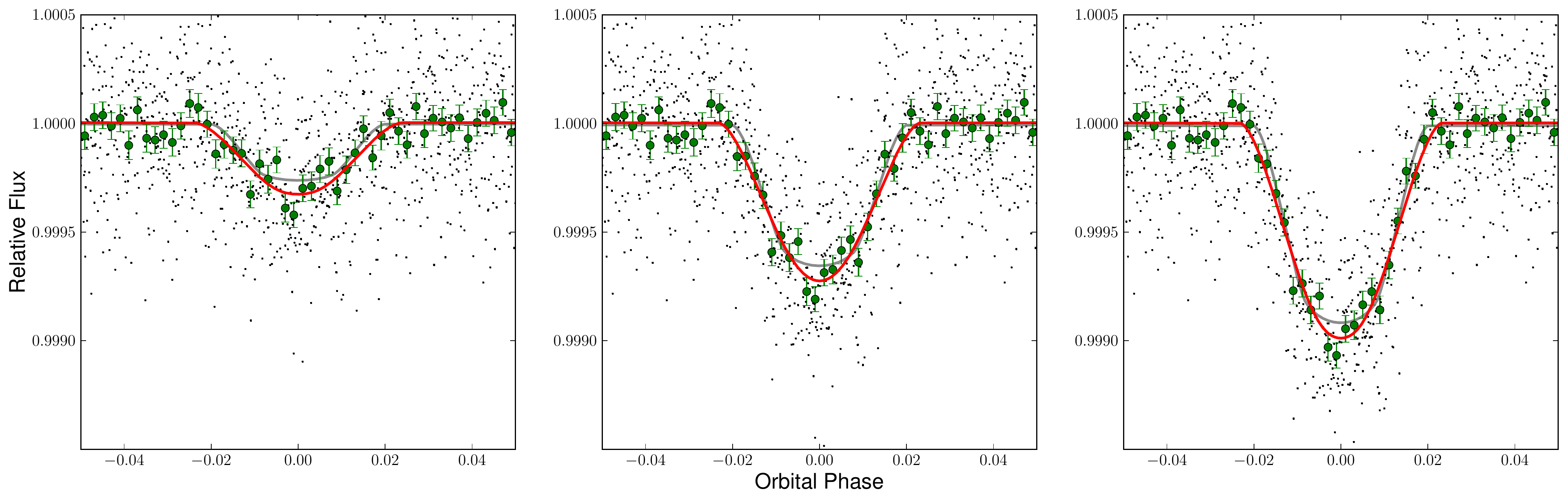}
\caption{The secondary and primary eclipses (upper and lower panels, respectively) of the BEB synthetic data with mass ratio $q = 0.3$, and impact parameter $b = 0.5$ for the three levels of dilution used for the simulations. The black dots are the individual binned data points (see text), the green circles are the average of the data in 0.0025-size bins. The red curve is the best fit model found using the MCMC algorithm and the gray curve is the actual model injected in the \kepler\ light curve. It can be seen that the presence of correlated data near the centre of the primary eclipse leads the fit procedure to a longer, more V-shaped eclipse. This highlights the importance of using realistic error distributions for the simulations.}
\label{fig.secondaries}
\end{figure*}

\begin{figure*}
\begin{center}
\includegraphics[height = 0.366\textwidth]{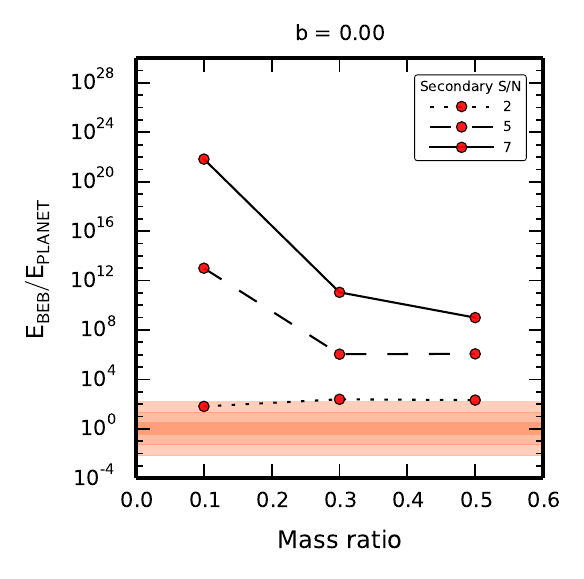}
\includegraphics[height = 0.366\textwidth]{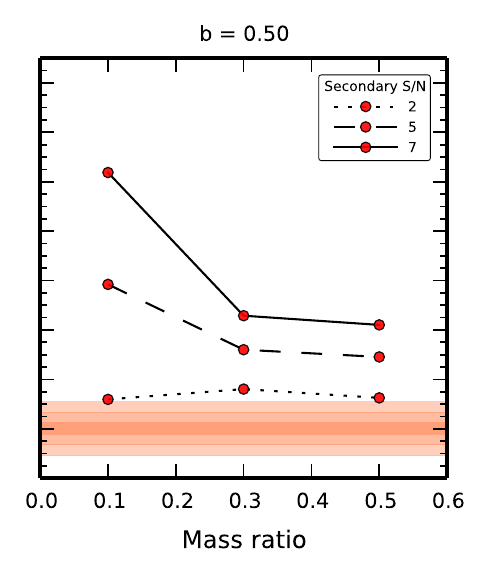}
\includegraphics[height = 0.366\textwidth]{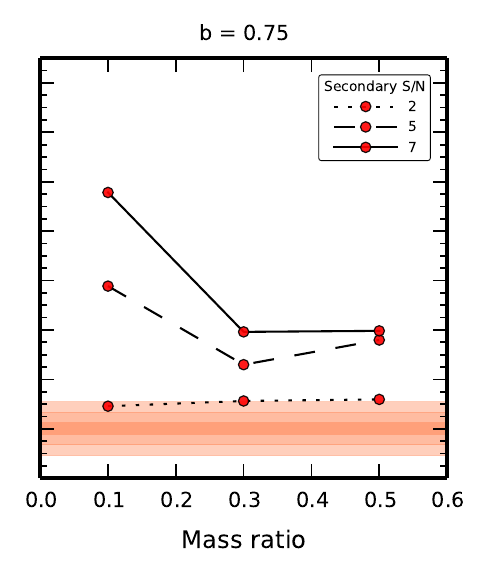}
\end{center}
\caption{Evidence ratio (Bayes factor) between BEB and PLANET models for the synthetic BEB data with impact parameter b = 0.0 (left), b = 0.5 (center), and b = 0.75 (right). In each panel, the results are plotted as a function of the simulated mass ratio of the diluted binary for different signal-to-noise ratios of the secondary eclipse: 2 (dotted curve), 5 (dashed curve), 7 (solid curve).  The 68.3\% confidence interval is smaller than the size of the symbols.} The shaded areas indicate the regions where the support of one model over the other is arbitrarily considered "Inconclusive" (dark), "Positive", "Strong" (light orange), and "Very strong" (white), according to \citet{kassraftery1995}.
\label{fig.resultsBEB}
\end{figure*}
 
\begin{comment}
\begin{figure*}
\begin{center}
\includegraphics[height = 0.38\textwidth]{plots/EvidenceRatio_BEB_b000_snr_vRev2_683confidence}
\includegraphics[height = 0.38\textwidth]{plots/EvidenceRatio_BEB_b050_snr_vRev2_683confidence}
\includegraphics[height = 0.38\textwidth]{plots/EvidenceRatio_BEB_b075_snr_vRev2_683confidence}
\end{center}
\caption{Same as Fig.~\ref{fig.resultsBEB} as a function of S/N of the secondary eclipse. In each figure, the results are presented for the three simulated mass-ratios, shown in different colors.}
\label{fig.resultsBEBsnr}
\end{figure*}
\end{comment}

\subsection{BEB simulations \label{sect.BEBresults}}

The BEB model consists of an eclipsing binary (EB) in the background of a bright star that dilutes its eclipses. The primary of the EB was chosen, as the primary of the planet scenario, to be a solar twin. Different values of the impact parameter, the binary mass ratio and the dilution of the eclipses were adopted for the synthetic data. The dilution of the EB light curve is quantified using the S/N of the secondary eclipse, measured using equation \ref{eq.snr}. The S/N of the diluted secondary eclipse in our simulations is 2, 5, or 7. These values were chosen to go from virtually undetectable secondaries to clearly detectable ones. For example, the \kepler\ pipeline requires S/N to be above 7.1 for a signal to be considered as detected \citep[e.g.][]{fressin2013}. In total, we produced light curves of 27 BEBs. An example of a secondary eclipse with the three levels of dilution can be seen in Figure~\ref{fig.secondaries}. The corresponding primary eclipse is also shown in the same Figure. 

Evidently the S/N of the primary eclipse changes as well with the dilution level, ranging from 53 to 370 (Table~\ref{table.resultsBEB}). In general, the primary eclipse S/N increases with S/N of the secondary, and diminishes with mass ratio $q$ and impact parameter $b$, as expected.

The Bayes factors in favour of the BEB model, $B_\mathrm{BP} = \prob{D}{H_\mathrm{BEB},I}/\prob{D}{H_\mathrm{PLANET},I} = E_\mathrm{BEB}/E_\mathrm{PLANET}$, are listed in Table~\ref{table.resultsBEB}, where the first four columns present the parameters of the simulated BEBs (including the primary eclipse S/N), and the following ones are similar to the columns in Table~\ref{table.resultsPLA}. The results are plotted as a function of the mass ratio $q$ of the simulated eclipsing binary in Figure~\ref{fig.resultsBEB}. The Bayes factor increases  with the S/N of the secondary eclipse, i.e. it \emph{decreases} with the dilution of the light curve. For the two lowest dilution levels, $B_\mathrm{BP}$ decreases with the mass ratio of the EB. This seems counterintuitive, since the ingress and egress times of the  eclipses become longer as $q$ increases, and therefore more difficult to fit with the PLANET model. The Bayes factor in favour of the BEB should therefore increase with $q$. However, as mentioned above, the primary S/N decreases as well towards bigger stars, which must counteract and dominate over this effect. Indeed, when the obtained Bayes factor is normalized by the S/N of the primary eclipse an inverse trend is seen with $q$. This means that the effect of the size of the secondary component of the EB is less important than that of the S/N of the primary eclipse. For secondary S/N = 2, the S/N of the primary changes less with $q$ and the size of the secondary component becomes the dominant factor. An inverse behaviour  with $q$ is then observed. Additionally, because the primary eclipse S/N does not change much with $b$, $B_\mathrm{BP}$ is approximately constant with impact parameter.

 As expected, for none of the simulations the Bayes factor strongly favors the PLANET hypothesis. Additionally, except for the highest dilution level and small mass ratio, all BEB scenarios are correctly identified by the data. This is because the PLANET model requires a large, evolved star to reproduce the shape and duration of the stellar eclipses, which is severely punished by the solar priors imposed on the target star. Nevertheless, all planet fits result in a stellar density $\rho_* < 0.28\,\rho_\odot$. The data clearly prefer this solution to one that would produce a worse fit but be more in agreement with the solar density prior (see Table~\ref{table.paramspriors}).
  
In table~\ref{table.resultsBEB},  the reduced $\chi^2$ of the best-fit model of each scenario is listed. As in the PLANET case, the fits obtained with both models are compatible with the dataset: a $\chi^2$  test fails to reject any of the obtained fit (see Sect.~\ref{sect.frequentist}).

\subsection{Including the prior odds \label{sect.inclHypPriors}}

The complete computation of the odds ratio requires specifying the ratio of the prior probabilities of the competing hypothesis, the prior odds. This ratio appears on the first term in the right-hand side of Eq.~\ref{eq.oddsratio}. As mentioned above, it depends on the environment of the target, the statistics of multiple stellar systems, planet occurrence rates, and the general structure of the Galaxy.

To compute the prior odds for our simulations, we assumed the same environment and follow-up observations that Kepler-22 \citep{borucki2012}, which include adaptive optics photometry (AO), speckle imaging and Spitzer observations. We refer the reader to this article for details about the available observations. For our simulations, we only considered the AO contrast curves obtained by \citet{borucki2012} in the J band. Additionally, we assumed that the simulated systems have the same apparent magnitude and position in the sky that Kepler-22. The Besan\c con galactic model was used to simulate a field around the target, and the AO contrast curve was used to compute the probability that a star of a certain magnitude lies at a given distance of the target without being detected. The stellar binary properties and occurrence rate for the simulated background stars were taken from \citet[][and references therein]{raghavan2010}. The 	planetary statistics were obtained from \citet{fressin2013}. 

The prior odds depend also on the observed planet size, which changes both the planet occurrence and the number of diluted binaries that are capable of reproducing the depth of the transit. The planetary radius of the simulated transit light curves was obtained from the transit depths assuming a 1-R$_\odot$ host star. The simulated planets correspond to the Earth (1 R$_\oplus$), Large Neptune (4.4 R$_\oplus$), and Giant categories (7.8 and 11.2 R$_\oplus$) of \citet{fressin2013}. Most of the simulated BEBs mimic planets in the Small Neptune category, but those with $q = 0.3$ and $q = 0.5$ and secondary S/N = 2 exhibit transits corresponding to planets in the Super Earth category.

\begin{figure}
\includegraphics[width = \columnwidth]{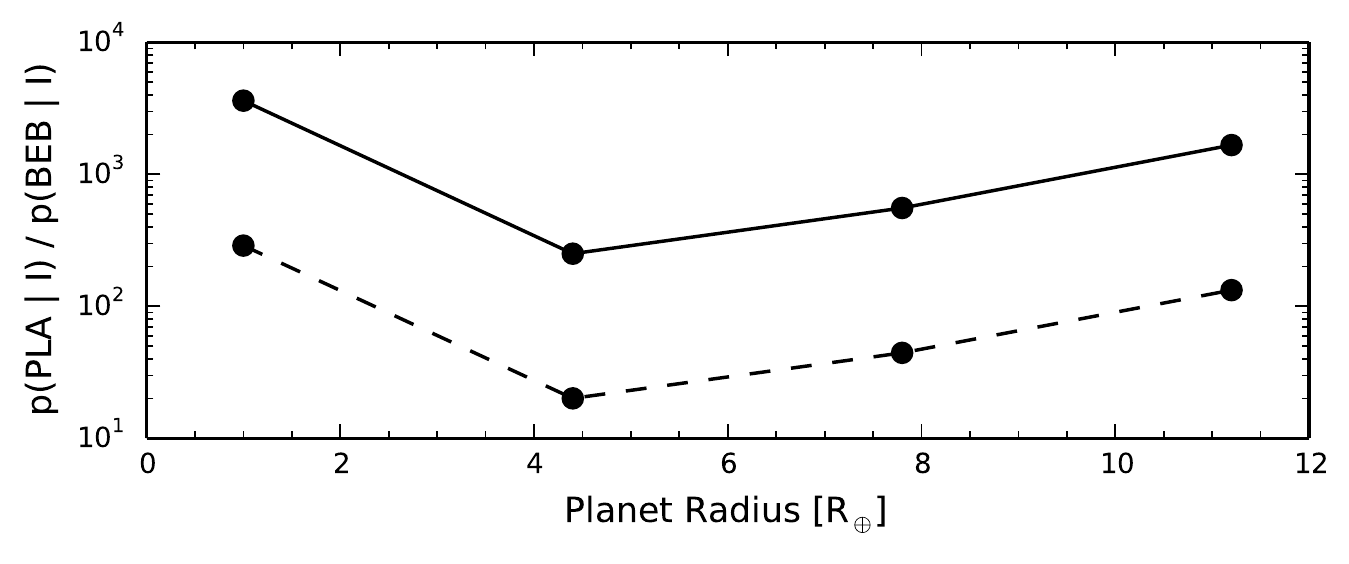}
\caption{Prior odds for the PLANET hypothesis as a function of planet radius (in logarithmic scale). The solid line is the value when the Kepler-22 contrast curve is used as constrain, and the dashed line represents a situation where no AO observation is available.\label{fig.hyppriors}}
\end{figure}

With all these assumptions, the prior odds can be computed as described in Sect.~\ref{sect.priors}. For the simulated systems the prior odds $\prob{H_{PLA}}{I}/\prob{H_{BEB}}{I}$ vary from around 250 for the 7.8-R$_\oplus$ planet to over 3600 for the Earth-size planet (Fig.~\ref{fig.hyppriors}). In the same figure we plot the prior odds obtained assuming no AO observations are available. In this case, we simply consider that the confusion radius is 2 arc seconds for stars 5 magnitudes fainter than the target star \citep[see][]{batalha2010}, and that it follows the same trend that AO contrast curve for other magnitude differences. This shows the value of precise AO observations, that drastically reduce the \emph{a priori} probability of having an unseen blended star in the vicinity of the target, and therefore equally reduce the prior probability of the BEB hypothesis. The uncertainties are estimated using a Monte Carlo method and are of the order of 10\%.

\begin{figure*}
\begin{center}
\includegraphics[height = 0.38\textwidth]{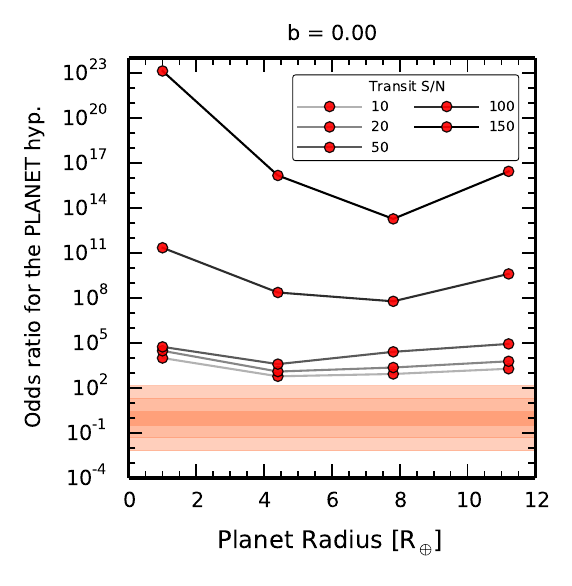}
\includegraphics[height = 0.38\textwidth]{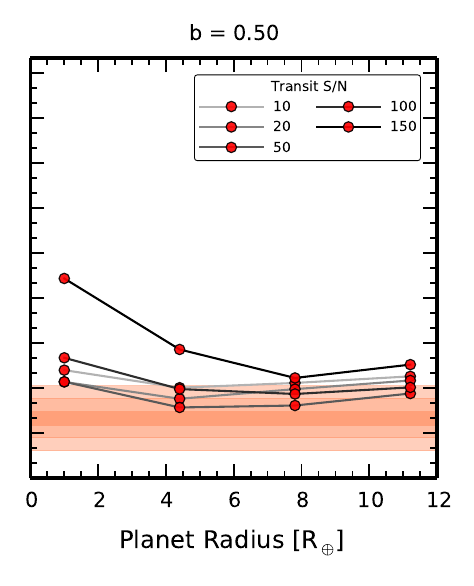}
\includegraphics[height = 0.38\textwidth]{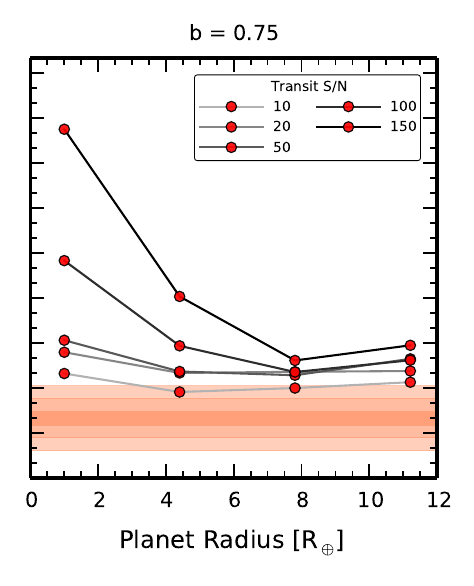}
\end{center}
\caption{Odds ratio in favour of the planet hypothesis as a function of planet radius for the synthetic planet data with impact parameter b = 0.0 (left), b = 0.5 (center), and b = 0.75 (right).  For the shaded areas see caption of Fig.~\ref{fig.resultsPLANET}.}
\label{fig.oddsratioPLA}
\end{figure*}

\begin{figure*}
%\begin{center}
\includegraphics[height = 0.37\textwidth]{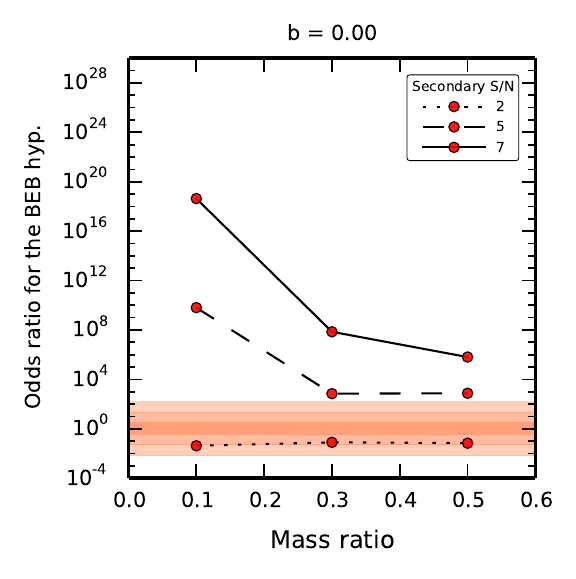}
\includegraphics[height = 0.37\textwidth]{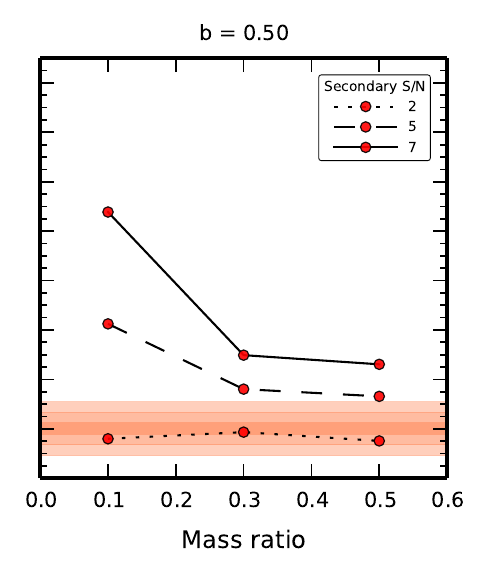}
\includegraphics[height = 0.37\textwidth]{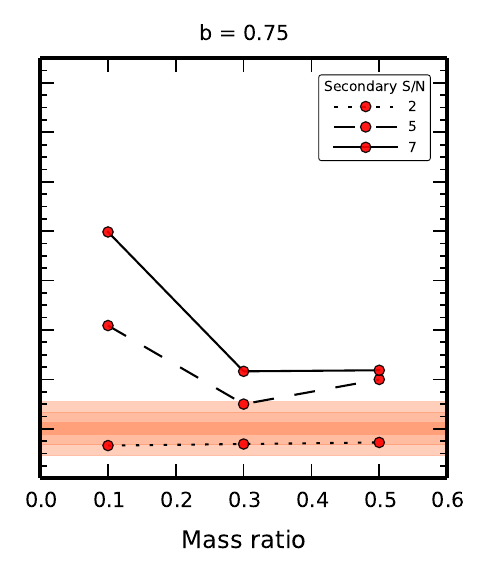}
%\end{center}
\caption{Odds ratio in favour of the BEB hypothesis as a function of binary mass ratio for the synthetic BEB data with impact parameter b = 0.0 (left), b = 0.5 (center), and b = 0.75 (right).  For the shaded areas see caption of Fig.~\ref{fig.resultsPLANET}.}
\label{fig.oddsratioBEB}
\end{figure*}

With these elements, the odds ratio is readily computed by multiplying the Bayes factors by the prior odds for the corresponding planet size. The results are listed in Tables~\ref{table.resultsPLA} and \ref{table.resultsBEB} and plotted in Figures~\ref{fig.oddsratioPLA} and \ref{fig.oddsratioBEB}. For the simulated planet light curves, the inclusion of the prior odds brings the odds ratio above 150 for almost all transit parameter sets. Even transits with S/N as low as 10 are now securely identified as planets. The odds ratio for low S/N curves (10 and 20) is strongly dominated by the prior odds. Therefore, the curves closely resemble those presented in Fig.~\ref{fig.hyppriors}. The exception remains the simulations at $b = 0.5$, stressing the importance of a more sophisticated error model. For the BEB simulations, the effect of including the prior odds is to \emph{diminish} the confidence of the identification based on the Bayes factor. The low probability of a blended eclipsing binary produces that some BEB scenarios cannot be identified as such, even if supported by the data. This is specially the case for systems with a high level of dilution (secondary S/N = 2).

\subsection{Other false positive scenarios} \label{sect.otherFP}

Other potential false positive scenarios described in Section~\ref{sect.fp} were studied in a less systematic way as done for the BEB scenario. We chose five synthetic Earth-size transit light curves and fitted them using a hierarchical triple model (TRIPLE), a background transiting planet model (BTP), and a planetary object in a wide binary model (PiB). The procedure was the same as above, and the results are synthesized in Table~\ref{table.resultsotherFP}. 

The hierarchical-triple scenario is easily rejected by the data in all cases, even for the lowest-S/N transits studied. In this model the EB is bound to the target star, which fixes the dilution level for a given set of stellar masses. Additionally, the radius ratio of the EB is limited by the stellar tracks and the age and metallicity of the target star, and the constrain that the dominant source of flux in the system be the target star. As a consequence, the ingress and egress times of the triple model are too long and do not fit the Earth-size transit correctly. Even allowing the target star to become much brighter than what the priors would allow does not improve the fit. This fact has been observed in various BLENDER validation cases \citep[e.g.][]{torres2011}. On the other hand, bigger planets should be better fitted using the TRIPLE scenario, since their ingress/egress times are longer.

Two facts limit the flexibility of the BEB and TRIPLE false positive scenarios and ultimately lead to their being rejected with respect to the PLANET model: the existence of a minimum stellar radius and the rareness of big, massive stars. The former is given by the limit in the stellar evolution tracks used to model stellar objects. The limit of the Dartmouth stellar tracks is 0.1 \msol\ (Table \ref{table.tracks}), which is close to the hydrogen-burning limit at around 80 \MJ. Most of the BEB and TRIPLE models trying to fit planetary transits tend to decrease the size of the eclipsing star as much as possible, and reach this limit. The second limitation is introduced in PASTIS through the priors of the stellar masses (Table \ref{table.paramspriors}).

False positive scenarios involving a transiting planet whose light curve is diluted by the presence of a second star do not suffer from the same limitations because the radius of the transiting object  can be reduced practically without limits. In this way, the BTP and PiB scenarios can mimic  the signal of an undiluted planetary transit well, and as a consequence, they cannot be correctly identified based on the light curve alone, as shown in Table~\ref{table.resultsotherFP}. This fact has already been reported by  \citet{torres2011} for much lower S/N transits. These authors note that in general it is possible to approximately reproduce the transit light curve of a planetary object of radius $R_p$ by a diluted system where both stars have the same brightness and the transiting planet has a radius larger by a factor $\sqrt{2}$. We show here that even in the case of very high-S/N transit light curves, additional observations are in general needed to reject these scenarios (see Sect.~\ref{sect.rv}). In addition, because the planet host star in the PiB scenario is bound to the target star, its prior probability is roughly of the same order than the planet hypothesis, and AO observations cannot reduce it significantly.

\begin{table}
\center 
\caption{Results of fitting other false positive scenarios to synthetic Earth-size transit data. \label{table.resultsotherFP}}
\begin{tabular}{ccc|c|ccc}
\hline
\hline
\multicolumn{3}{c|}{Model} & Scenario & \multicolumn{3}{c}{Bayes Factor} \\
b & snr & Rpl  & & $\log_{10}B_\mathrm{P;FP}$ &$\sigma_{+}$ & $\sigma_{-}$  \\
\hline
0.0 & 150 & 1.0 & TRIPLE & 21.63 & 0.11 & 0.14\\
	&	&	& BTP	& -0.813 & 0.051 & 0.084\\
	&	&	& PiB	& -0.491 & 0.045 & 0.077\\
\hline
0.5 & 150 & 1.0 & TRIPLE & 9.99 & 0.14 &  0.20\\
	&	&	& BTP	& -0.85 & 0.15 & 0.17\\
	&	&	& PiB	& -1.23 & 0.03 & 0.12\\
\hline
0.0 & 100 & 1.0 & TRIPLE & 9.55 & 0.13 & 0.18\\
	&	&	& BTP	& -0.72 & 0.09 & 0.12\\
	&	&	& PiB	& -0.95 & 0.06 & 0.11\\
\hline
0.0 & 20   & 1.0 & TRIPLE & 30.83 & 0.56 & 0.58\\
	&	&	& BTP	& 0.69 & 0.08 & 0.08\\
	&	&	& PiB	& 0.04 & 0.18 & 0.16\\
\hline
0.0 & 10   & 1.0 & TRIPLE & 18.58 & 0.09 & 0.16\\
	&	&	& BTP	& 0.48 & 0.05 & 0.07\\
	&	&	& PiB	& 2.99 & 0.09 & 0.14\\

\hline 
\end{tabular}
\end{table}

\section{Discussion} \label{sect.discussion}

The results from the previous section show that PASTIS is able to validate planets and to correctly identify false positives based on the analysis of light curve data alone. However, we find that only when the signal is large enough do light curves alone strongly favor one model over the other, and this not even for all false positive scenarios. In particular, false positive scenarios involving a transiting (giant) planet system, whose light curve is diluted by the presence of a second star (either in the system or aligned with it) seem to be able to mimic small-planet transits very precisely, and cannot be rejected by the light curve data alone, even in the high-S/N regime we have explored. Strong reliance on the priors odds is then required to validate transiting planets against these scenarios. On the other hand, the hierarchical-triple system scenario is unable to reproduce the transits of small-size planets. Background eclipsing binaries are correctly identified as such when the secondary eclipse of a diluted eclipsing binary has a S/N above around 5. In the particular cases simulated here, the out-of-transit variation does not seem to contribute to the correct identification of BEBs: no significant changes in the Bayes factor are observed when only the eclipses are fitted. Of course, this will depend on the orbital period of the system, and it is to be expected that reflected light, ellipsoidal modulation, and Doppler "boosting" variations \citep[e.g.][]{faigler2012} would be relevant for shorter-period candidates. Concerning transiting planets, our synthetic light curve data conclusively support the correct model if the transit S/N is higher than about 50 - 100 (for central transits; see Fig.~\ref{fig.resultsPLANET}). For the lower S/N the correct identification rests instead on the priors odds. Assuming typical conditions and follow-up observations of a target in the \kepler\ field, transiting planets are correctly recognized down to transit S/N = 10.  Of course, additional follow-up observations could provide additional support to validate the low-S/N transit signals without depending as much on the prior odds. We investigate this briefly in Sect.~\ref{sect.rv}.

\subsection{Implication for the validation of \kepler\ candidates} \label{sect.koivalidation}

\begin{figure}
\includegraphics[width = \columnwidth]{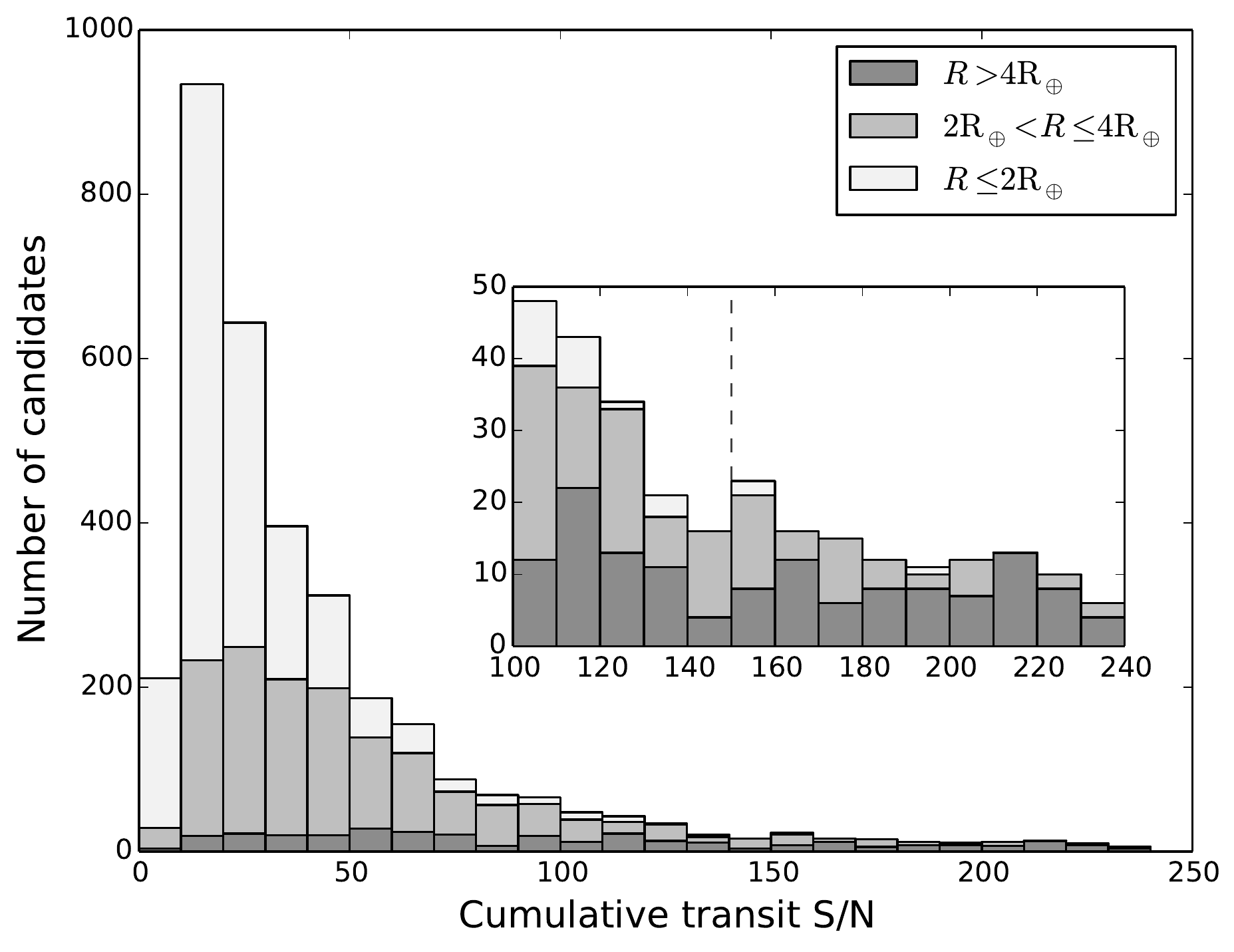}
\caption{Transit signal-to-noise ratio (S/N) for all \kepler\ objects of interest (KOI). The S/N is computed over all quarters of available data. The inset shows the range between S/N 100 and 240. Only five KOIs with estimated radius below 2 $R_\oplus$ have transit S/N above 150. The 3592 KOIs marked as CANDIDATES in the NExScI table are included.}
\label{fig.snhist}
\end{figure}

Transits of Neptune-size and Jupiter-size objects are observed with S/N well above 150 by the \kepler\ mission, except for the faintest target stars or the longest-period candidates for which a reduced number of transits have been observed. On the other hand, only 168 (respectively 67) \kepler\ candidates with estimated radius below 4 R$_\oplus$ have S/N above 100 (respectively 150). These numbers are reduced to 25 and 5 for candidates smaller than 2 R$_\oplus$\footnote{The five candidates with S/N above 150 are: KOI-69.01, KOI-70.02 (Kepler-20 b), KOI-72.01 (Kepler-10 b), KOI-245.01 (Kepler-37 d), KOI-268.01 }. No candidate with estimated radius below 1.4 R$_\oplus$ has S/N above 150, and only five have S/N above 100\footnote{Kepler-10 b, KOI-82.02, KOI-85.02 (Kepler-65 b), KOI-1300.01, KOI-1937.01}. In Fig.~\ref{fig.snhist} we present the histogram of the cumulated S/N of all the \kepler\ candidates\footnote{Data was obtained from the NExScI: \url{http://nexsci.caltech.edu/ }}. Additionally, our results assume that the cadence of the observations is roughly 1 minute. This is not true for most of the \kepler\ targets, which are measured on a 30-minute cadence. In these cases, the light curves are smeared and the resolution of the ingress and egress phases is reduced. This should mainly affect our results for small-size planets (see Sect.~\ref{sect.caveats}).

The vast majority of \kepler\ candidates, then, cannot be validated by this method by studying their light curve alone. Strong reliance on additional data and on the hypotheses prior odds seems to be the ineluctable. The BLENDER validations \citep[e.g.][]{fressin2011, fressin2012, borucki2012} exemplify this fact. On the other hand, over 360 \kepler\ planet candidates have S/N above 150. Among them, there are all the unresolved cases from \citet{santerne2012}. Since \citet{santerne2012} focused on giant-planet candidates, the scenarios involving diluted planetary companions should be easily discarded, and the candidates could be promptly confirmed using the \kepler\ light curve alone. This is outside the scope of the present paper and is deferred to a follow-up article.

\subsection{Contribution from RV observations} \label{sect.rv}

Radial velocity and absolute photometry can help constrain false-positive scenarios, and tip the scale for the cases where the light curve alone is not enough to decide between different scenarios. A detailed study of the contribution of RV observations will be given elsewhere (Santerne et al., in prep.). Here, we study the issue by considering some of the false positive hypotheses simulated above. We computed the RV signal of the maximum-posterior model for each parameter set at the quadratures of the orbit, which are usually the first orbital phases observed when performing ground-based follow-up. Additionally, we computed the signal that would be observed if the orbit is randomly sampled in 10 or 25 phases, since, depending on the relative center-of-mass velocities of the objects, the maximum RV signal (including also the maximum bisector, contrast, and width variations) can occur at any moment of the orbit.

For the BEB and BTP scenarios, we randomly assigned a center-of-mass velocity for the background system and the target star by drawing samples from the \citet{nordstrom2004} distribution of radial velocities of nearby stars. The rotational velocities of all stars were fixed to $v \sin i_\star = 4$ \kms. The process was repeated 200 times for each parameter set, and the observed amplitude was recorded. We then computed what fraction of the sample exhibited an amplitude of the RV and bisector velocity span larger than a certain value.

Only a few of the BEB models to synthetic BEB light curves exhibit RV amplitudes larger than 2.5 \ms, and none exhibits variations above 25 \ms. This is expected given the level of dilution produced by the target star. Indeed, the brightness differences between the target star and the background binary are larger than 5 magnitudes in the Johnson-V band for these models, which is too large to expect substantial signal in any of these observables (see Santerne et al. in preparation). A similar situation is found for the bisector velocity span of the CCF. Eight BEB models include binaries that are between 3.0 and 5.0 magnitudes fainter than the target star. They exhibit RV amplitudes above 2.5 \ms\ for most of the sampled center-of-mass velocity differences, but for none of them the RV amplitude is above 10 \ms. In a few BEB models the brightness difference between the binary and the target star is less than 3 magnitudes. These systems show RV variations with amplitudes above 10 \ms\ in around half of the sampled velocity differences. All these systems have secondary eclipses with S/N 5 or 7, and are therefore clearly identified from the light curve analysis. We list them in Table~\ref{table.BEB_rvcontrib}, together with some examples of more strongly diluted BEBs. We conclude that radial velocity observations do not help in pinpointing the BEB scenarios that cannot be identified from the light curve data alone.

\begin{table}
%\center
\caption{Percentage of cases in which a RV signal larger than a given amplitude will be observed if 25 data points sample the orbital phase for selected simulated BEB systems (see text). \label{table.BEB_rvcontrib}}
\begin{center}
\begin{tabular}{ccc|c|lll}
\hline
\hline
\multicolumn{3}{c|}{Model}  & & \multicolumn{3}{c}{Amplitude [\ms]} \\ 
b	& q	& snr & $\Delta\,m_V^\dagger$ & $>2.5$ & $>10$ & $>25$ \\
\hline
0.00 & 0.1 & 7 & 3.0 & 65.0 & 46.0 & 00.0  \\
0.50 & 0.1 & 7 & 2.6 & 64.5 & 57.5 & 00.0 \\
0.75 & 0.1 & 5 & 2.6 & 59.0 & 53.5 & 00.0 \\
0.75 & 0.1 & 7 & 2.7 & 61.5 & 37.0 & 00.0  \\
0.00 & 0.1 & 2 & 4.6 & 35.0 & 00.0 & 00.0	\\
0.50 & 0.1 & 2 & 4.3 & 74.0 & 00.0 & 00.0\\
0.75 & 0.1 & 2 & 4.3 & 57.0 & 00.0 & 00.0\\
0.00 & 0.3 & 2 & 6.8 &  00.0 & 00.0 & 00.0\\
0.00 & 0.5 & 2 & 7.7 &  00.0 & 00.0 & 00.0\\
\hline
\end{tabular}
\end{center}
\small{$\dagger$} Apparent magnitude difference in the V band between the target star and the diluted binary.
\end{table}

\begin{figure}
\center
%Printer
%\includegraphics[width = 0.33\textwidth]{plots/RVcontrib_PLANET_median_b000_v3}
%\includegraphics[width = 0.33\textwidth]{plots/RVcontrib_PLANET_median_b050_v3}
%\includegraphics[width = 0.33\textwidth]{plots/RVcontrib_PLANET_median_b075_v3}
\includegraphics[width = 0.36\textwidth]{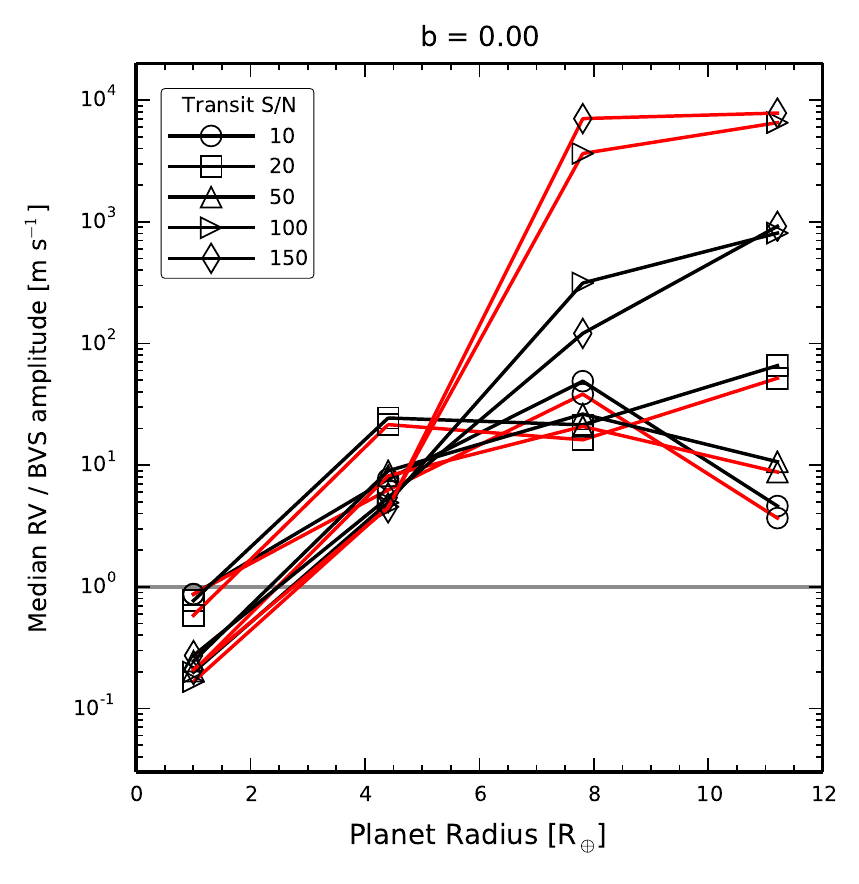}
\includegraphics[width = 0.36\textwidth]{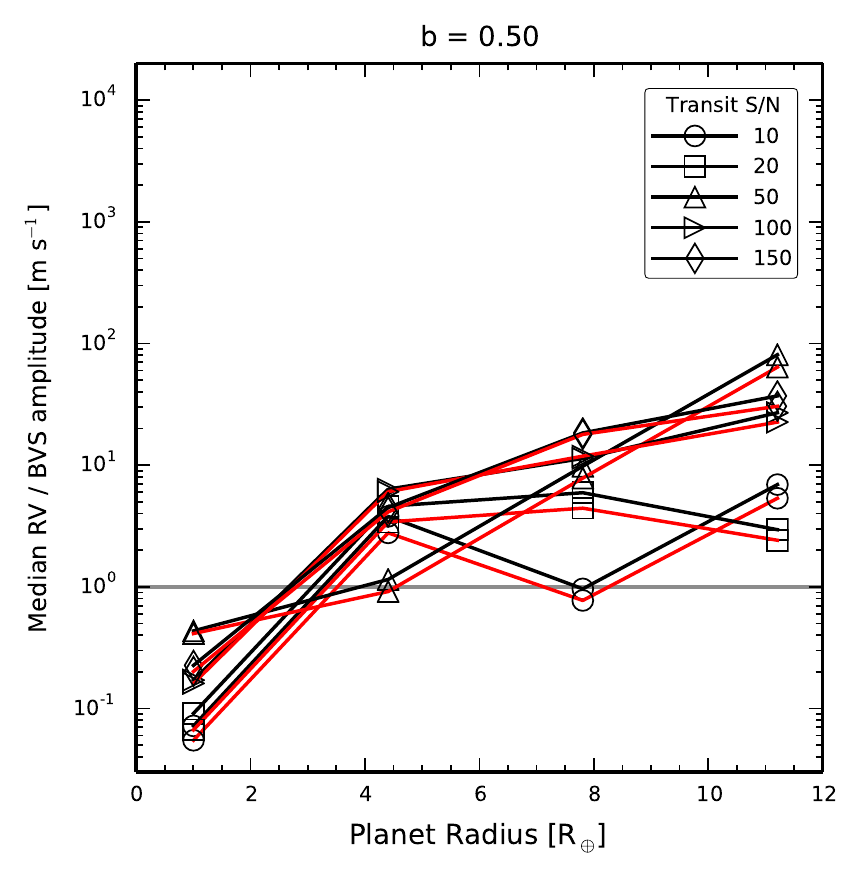}
\includegraphics[width = 0.36\textwidth]{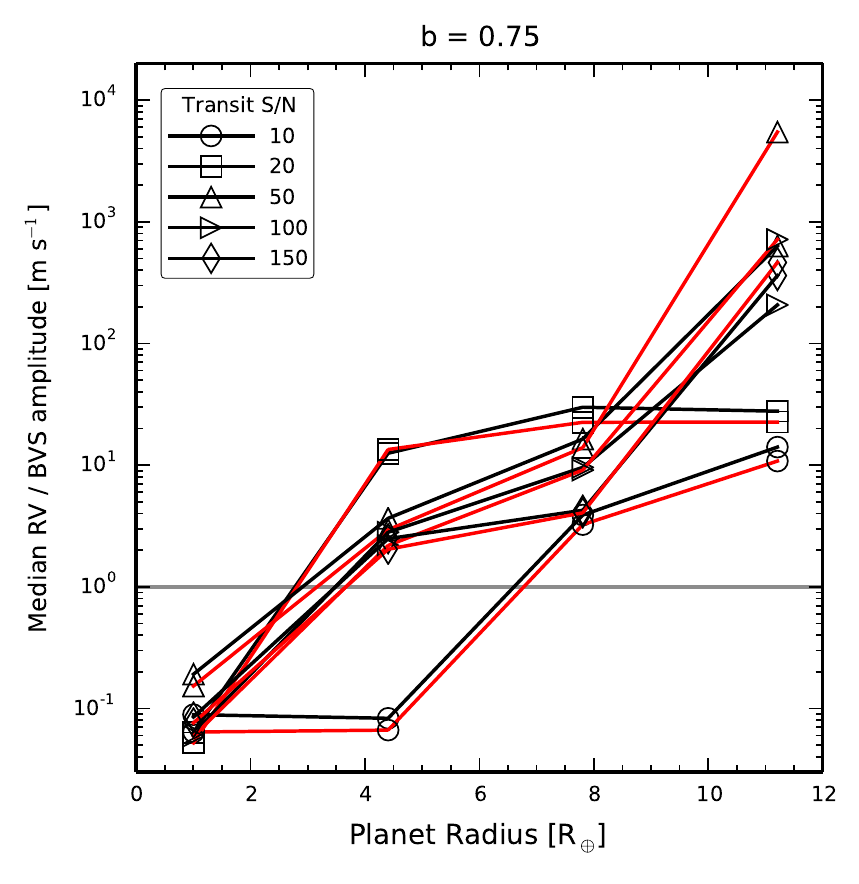}

%Referee
%\includegraphics[width = 0.7\columnwidth]{plots/RVcontrib_PLANET_median_b000_v2}
%\includegraphics[width = 0.7\columnwidth]{plots/RVcontrib_PLANET_median_b050_v2}
%\includegraphics[width = 0.7\columnwidth]{plots/RVcontrib_PLANET_median_b075_v2}
\caption{Median expected RV (black) and BVS (red) amplitude of the maximum-posterior BEB scenarios as a function of the planetary radius of the synthetic light curve, based on 200 randomly drawn relative radial velocities (see text).  The orbital phase of the eclipsing binary is sampled randomly using ten points.  The horizontal line indicates the 1 \ms\ level. The saturation seen for the largest planet scenarios at $b = 0.0$ is due to the magnitude condition mentioned in Sect.~\ref{sect.synthLC}. \label{fig.rvcontribPLA}}
\end{figure}

On the other hand, the maximum-posterior BEB models to synthetic planet data, in principle, produce detectable RV amplitudes. Failure to detect these variations in available velocimetric data could help discarding a BEB as a viable hypothesis. Following the same procedure as above, we find that some of the BEB maximum-posterior models produce a detectable signal both in the radial velocity amplitude and in the bisector velocity span, specially for large planet candidates. In Fig.~\ref{fig.rvcontribPLA} we show the median amplitude obtained using ten points to sample the orbit. It can be seen that the amplitude increases for bigger planets, reaching values around 100 \ms, and sometimes above 1 \kms, even for low-S/N transits. The trend is opposite to the one observed in Fig.~\ref{fig.resultsPLANET}, which implies that radial velocity data is helpful in the analysis of the transiting planets that cannot be identified using only the light curve. These amplitudes are explained by the magnitude difference between the target star and the background binary system. Except for the Earth-size simulations, all the maximum-posterior BEB models have magnitude differences smaller than 5 in the V-band. For the simulations with 7.8 R$_\oplus$ and 11.2 R$_\oplus$ the difference is usually smaller than two magnitudes. In these cases, we also expect absolute photometry (SED fitting) or spectroscopic data to provide additional constraints.

The BTP model does not exhibit RV variations above 2.5 \ms\ for any of the planet scenarios explored. As in the BEB simulations, this is due to the high level of dilution needed to reproduce Earth-size transits.

Finally, the RV signal produced by the maximum-posterior PiB scenarios should also be detectable. Indeed, when sampling the orbit with ten randomly distributed measurements, the median amplitude of the RV and signal is between 20 \ms and 30 \ms. The median bisector amplitude, on the other hand, is smaller than 2 \ms.

\begin{comment}
\begin{table}
%\center
\caption{Percentage of cases in which a RV signal larger than a given amplitude will be observed if 25 data points sample the orbital phase for the BTP maximum-posterior model. \label{table.BTP_rvcontrib}}
\begin{center}
\begin{tabular}{ccc|llll}
\hline
\hline
\multicolumn{3}{c|}{Model}  & \multicolumn{3}{c}{Amplitude [\ms]} \\ 
b	& Rpl& snr  & $>2.5$ & $>10$ & $>25$ & $>50$\\
\hline
0.00 & 1.0 & 100 &  43.5 & 41.5 & 38.5 & 0.5 \\
0.00 & 1.0 & 150 &  44.5 & 42.0 & 16.0 & 1.5 \\
0.50 & 1.0 & 150 &  36.5 & 28.5 & 21.5 & 4.5 \\
\hline
\end{tabular}
\end{center}
\end{table}
\end{comment}

\subsection{Comparison with the frequentist approach.} \label{sect.frequentist}
It is interesting to compare the results from the Bayesian model comparison to what would be obtained by a frequentist hypothesis testing method. The main difficulty faced by frequentist hypothesis testing is that the ranking of models cannot be rigorously made: either a hypothesis is rejected or nothing else can be said about it, the concept of the probability of propositions being completely alien to the frequentist. None of the fits obtained during our tests, whether using the correct model or not, can be rejected based on a $\chi^2$ test. Indeed, the values of the reduced $\chi^2$ statistics are below 1 for all models (see Tables~\ref{table.resultsPLA} and \ref{table.resultsBEB}). All the fits are reasonably "good".

A series of workarounds to overcome this limitation exists. For example, the F-test permits comparing the reduced $\chi^2$ of two competing fits to the same dataset, but it is known to be extremely sensitive to the assumption of normality of the error distribution. The likelihood-ratio test is also a possible way to compare models, but it only works for nested models (i.e. pairs of models where the simplest one is equal to the more complex one with constraints on one or many parameters).

Disregarding these conditions, we nevertheless applied these two tests to our results. In Tables~\ref{table.resultsPLA} and \ref{table.resultsBEB} we list the F-test probability and the likelihood-ratio test statistics $D = - 2 \ln(\mathcal{L}_0/\mathcal{L}_1)$, where $\mathcal{L}_0$ and $\mathcal{L}_1$ represent the maximum likelihood value of the null and alternative hypotheses, respectively. For the likelihood-ratio test we took the PLANET model as the null (simplest) hypothesis. We also list the p-value of $D$ (for $D > 0$), which was computed by approximating the distribution of $D$ as a $\chi^2$ distribution with number of degrees of freedom equal to the difference between the number of parameters of each model. For the planetary synthetic data, we found that the F-test is not capable of distinguishing between both models, the \emph{p}-values being above 23\% for all parameter sets. Similarly, the likelihood-ratio test does not allow rejecting the null hypotheses over the more complex BEB model. For some cases, specially those with central, high-S/N transits, the PLANET model even fits better than the BEB model. The F-test does not perform better for the BEB synthetic data: the \emph{p}-values are above 0.27 for all simulated contaminated binaries. On the other hand, the likelihood-ratio test allows discarding the PLANET model for practically all tested parameters, in some cases with a very high significance. Note that the $D$ statistics is well correlated with the S/N of the \emph{primary} eclipse, which are all above 50 for these examples. The largest \emph{p}-value is 7.7\%, for a system with primary S/N = 65. It is to be expected that the $D$ statistics would fail to reject the PLANET model for systems with an even higher level of dilution, where the primary eclipse would have S/N $< 50$, and the secondary eclipse would be completely absent. In those cases, one would expect the performances of our method to resemble those of the planetary cases.

Frequentists hypothesis testing methods are therefore less efficient in confirming the nature of the majority of the synthetic light curves used in Section~\ref{sect.application}, specially for the planetary cases. The reason is that these tests are based on point estimations of the best likelihood value. The relevant quantity in Bayesian approach, instead, is the hypothesis evidence (equation~\ref{eq.evidence}), which is the expectation value of the likelihood over the entire prior space. The evidence has information on the \emph{entire distribution} of the likelihood, and not only its maximum value. It provides, therefore, more information about the hypotheses being tested, and permits, in particular, to rigorously select one over the other.

Additionally, Bayesian analyses naturally separate the contribution of the data and the hypotheses priors to the odds ratio. This permits studying the weight of prior odds on the final outcome, as we have done, but also coming back to the same system with updated knowledge on the priors, without need to reanalyze the data.

\subsection{Comparison with previous validation methods.} \label{sect.blender}

Some features of the BLENDER tool and the method by \citetalias{morton2012} have already been mentioned in Sect.~\ref{sect.planetvalidation}. Here we describe in detail some of the most important differences between PASTIS and these two techniques.

Concerning the statistical formalism, the BLENDER approach relies on a grid of $\chi^2$ differences between a series of false-postive models and the planetary model, which is constant over the grid. The $\chi^2$ is computed using only the \kepler\ light curve. The critical values are taken from a $\chi^2$ distribution with number of degrees of freedom equal to the number of free parameters in the false positive model studied \citep{fressin2011}. By doing this, BLENDER uses a kind of likelihood-ratio test\footnote{For it to be an actual likelihood-ratio test, the number of degrees of freedom of the $\chi^2$ distribution used to compute the critical values should be the difference of degrees of freedom between both models.}, even if the models compared are not nested. Therefore, although intuitively the $\chi^2$ difference seems a reasonable way to quantify the relative merits of the models, the use of this statistics to perform a rigorous model comparison is not justified from a statistical point of view. \citetalias{morton2012}, on the other hand, obtains the Bayesian evidence of each competing hypothesis by computing the integral over the three parameters of the trapezoidal model employed to fit the light curve. A clear advantage of this approach is the low dimensionality of the problem, which allows direct integration. Because the model is non-physical, the likelihood distribution is \emph{exactly} the same for all hypotheses. Only the prior distribution changes from one hypothesis to the other, producing different values of the evidence. On the other hand, the simplicity of the model implies that not all information available in the data is being used, although the three parameters of the trapezoid model do contain the essential characteristics of a transit light curve, which may suffice for low-S/N transits, such as those detected by \kepler. In PASTIS, as much information as possible is taken into account by employing more complex, physically-motivated models of the data. This largely increases the dimensionality of the problem, with all the difficulties this implies, but should permit to conclude on candidates where the \citetalias{morton2012} technique cannot. For example, our simulations from the previous section show that, using data that will be characteristic of future space missions, our technique should validate candidates \emph{independently} of the prior distribution (or the populations of \citetalias{morton2012}).

BLENDER and the \citetalias{morton2012} technique share the way additional observations are taken into account: a given false positive model is either compatible with them or not at all. In BLENDER, the \kepler\ light curve is also used in this yes/no manner to eliminate false positives \citep[e.g.][]{torres2011}.This methodology raises the issue of where the limit shall be put. For example, in BLENDER, a 3-$\sigma$ limit is used for the stellar colors and the Spitzer transit depth; \citetalias{morton2012} uses 0.1 magnitudes as the maximum allowed difference in the stellar colors. The choice of these limits, although reasonable, remains arbitrary. Additionally, the false positives located outside these limits will usually be much more numerous than those inside the limits, which compensates (at least partly) their inadequacy to explain the observations. PASTIS treats complementary data and the transit light curve issued from discovering survey on equal footing. The philosophy is to model all data simultaneously. In this way we guarantee that the uncertainties of \emph{all} datasets are correctly propagated to the final odds ratio, and we do not need to mind about the limiting values at which false positives are discarded.  In fact, as the Bayesian approach presented here integrates over the entire parameter space, all false positives are considered, weighting each scenario, in a way, by its likelihood value.

All validation procedures use Galactic population syntheses models to compute the frequency of background false positives. BLENDER uses the Besan\c con Galactic  simulations together with a Monte Carlo procedure that compare synthetically-generated blends with the BLENDER constraints \citep[e.g.][]{borucki2013}. \citetalias{morton2012} uses a similar procedure but employing the TRILEGAL \citep{girardi2002} model. As mentioned in Sect.~\ref{sect.priors}, PASTIS also uses these models. In any case, since the observed color index is compared with that of simulated stars, the interstellar extinction employed is of crucial importance. Additionally, the number of possible blends depends as well on the extinction in the line of sight. To model the extinction the Besan\c con Galactic model uses, by default, a disc of diffuse absorbing matter, which is not suitable for low Galactic latitudes \citep{robin2003}. \citet{torres2011} show that extinction does not have a strong effect on their results, but they use a representative constant extinction coefficient. Similarly, \citetalias{morton2012} uses a calibration of the extinction at the infinity.  Some regions of the \kepler\ field being around $l =10$ degrees \citep[e.g.][]{ballard2011}, these techniques may suffer from an inadequately chosen interstellar extinction model. For PASTIS, we couple the Galactic models with a realistic three-dimensional model of the Galactic extinction, which reproduces correctly the distribution of stars obtained from observations (see Sect.~\ref{sect.priors}).

\subsection{Caveats and limitations} \label{sect.caveats}

The results of the analysis using synthetic light curves presented in Section~\ref{sect.application}, and some of the implications described in this Section depend on some implicit assumptions that we discuss here. 

In the first place, the synthetic signals have been injected in the reduced PDC \kepler\ light curve of KOI-189. At this stage of the \kepler\ pipeline, the data have gone through a series of corrections  \citep[see][and references therein]{jenkins2010, christiansen2013}, which, we assume, preserves the features of the injected signal identically. In reality, some level of distortion, albeit small, or suppression of features, is to be expected. This is specially the case for the diluted secondary eclipses, which have low S/N and are therefore more prone to be affected by the \kepler\ pipeline \citep{christiansen2013}. By neglecting this effect, we are probably overestimating the capacity  of our method to distinguish one model from the other, which relies on light curve features such as the duration of the ingress and egress phases, or the presence of  secondary eclipses and out-of-transit modulation.

Secondly, in our simulations, we have fixed the ephemerides of the transits, and set the eccentricity equal to zero (the correct value). In reality, eccentric orbits add to the complexity of the problem. For example, an eccentric BEB not exhibiting secondary eclipses could resemble more closely a planetary light curve than the same EB on a circular orbit. This should be kept in mind when interpreting our results. On the other hand, eccentric EB should be more easily identified from RV data because the amplitude is larger for a given $q$, and therefore there are increased chances that velocity of the binary components will coincide with that of the target star at a given point in the orbit, which would produce an effect in the bisector velocity span.  Concerning the ephemerides, the effect of fixing it is not expected to have an important effect on our results, because these parameters are usually well constrained. Additionally, scenarios where the period of the EB is twice the nominal period of the candidate seem in general unable to reproduce the transit of a small planet \citep{torres2011, fressin2011}. 

The simulated models were injected in real \kepler\ {\emph short cadence} (SC) data. SC data have a cadence resembling that of PLATO, which will have 32 "normal" telescopes read out with a cadence of 25 seconds \citep{rauer2013}. However, this should be taken into account when interpreting our results in the context of the \kepler\ mission, because most of the \kepler\ targets do not have this high sampling rate but rather have one point every about 30 minutes. This sampling rate smears out the transit curve and reduces the resolution of the ingress and egress phases, which are of fundamental importance to compare planetary and FP models. As a consequence, our simulations surely \emph{overestimate} the capabilities of PASTIS for general application to \kepler\ candidates. Of course, this issue is more stringent for small-size planets, for which the duration ingress/egress phases are comparable with the \kepler\ long-cadence data sampling rate. The results on giant planets should not be strongly affected by this assumption.

Our results depend strongly on the method used to estimate the Bayesian evidence (eq.~\ref{eq.evidence}). The TPM estimator by \citet{tuomijones2012} has the advantage of being easy to compute based on the posterior samples obtained with the MCMC. However, being insensitive to the size and shape of the parameter priors employed \citep{tuomijones2012}, TPM underestimates the penalization exercised by the Occam's factor on models with large number of parameters. As all false positive models have more parameters that the planet model, the reported Bayes factors for the PLANET model  are underestimated. On the one hand, this reduces the possibility of validating the planet hypothesis simply because the underlying model have fewer parameters than the competing false positive hypotheses. On the other hand, comparing models based on the values provided by the TPM estimate might not be statistically rigorous. We have recently started testing additional methods \citep[e.g.][]{chibjeliazkov2001} to study this potential issue.

We have shown that for most currently available small-size transiting candidates, statistical validation must rely heavily on the prior odds. However, our knowledge on the factors on which these priors depend is not without uncertainty, and correctly quantifying these uncertainties in the final odds ratio value does not seem straightforward. At its current state, this is perhaps the main limitation of the statistical validation procedure, independently of the procedure or approach taken. This is certainly bound to change in the future: Gaia should permit refining the Galactic population and reddening models, as well as improve our knowledge on giant planet populations and binary statistics, and PLATO should provide high-S/N transits of small-size candidates, which will permit validation based mostly on the data (see Sect.~\ref{sect.application}).

\section{Conclusions}\label{sect.conclusions}
At present, planet validation is the only technique capable of establishing the planetary nature of the smallest transiting candidates detected by the \emph{CoRoT} and \kepler\ space missions. The planetary hypothesis is compared with all possible false positives, and the planet is considered validated if it is found to be much more probable than all the others. Unless one of the competing hypotheses can be rejected as a possible explanation for the data, which is rarely the case, a rigorous comparison of the different hypotheses has to be made in a Bayesian framework. We have presented a method to self-consistently model most of data usually available on a given candidate --the discovery light curve, the radial velocity follow-up observations, light curves obtained in different photometric filters, absolute photometric observations of the target star-- under different competing hypotheses relevant to the problem of planet validation. Using these models, we compute the Bayesian odds ratio via the importance sampling technique. This procedure has been implemented in a \emph{python} package named PASTIS (Planet Analysis and Small Transit Investigation Software).

The posteriors of the model parameters are sampled with a MCMC algorithm. MCMC algorithms are much more efficient in sampling the posterior distribution of multidimensional problems than other more straightforward methods, such as grid evaluation. Therefore, we can use models with an arbitrary number of parameters. This allows us to add complexity to our models (such as limb-darkening parameters, or planetary and stellar albedos) at virtually no cost. Furthermore, the samples obtained with the MCMC algorithm are used to estimate the Bayesian evidence via importance sampling.  The MCMC algorithm implemented in PASTIS deals with parameter correlations by regularly performing a Principal Component Analysis, and takes into account the correlated nature of MCMC samples by thinning the chains using the measured correlation length. This method was shown to produce satisfactory results by comparing it to another existing MCMC code \citep[\emph{emcee};][]{emcee}.

The entire PASTIS planet-validation procedure was tested  using synthetic light curves of transiting planets and background eclipsing binaries (BEBs) whose eclipses are diluted by a brighter star. We separated the analysis in two parts, naturally present in Bayesian model comparison: a) the computation of the Bayes factor, which contains all the support the data give to one model over the other, and b) the computation of the odds ratio, which includes the prior odds, independent of the data.

For part a), we have found that the light curves of BEBs posing as transiting planets strongly support the BEB model if the dilution level is such that the secondary eclipse has S/N above about 5. Light curves with secondary eclipse S/N of 2, on the other hand, give marginal support for the correct model (see Fig.~\ref{fig.resultsBEB}). The dependence on the mass ratio $q$ seen in Fig.~\ref{fig.resultsBEB} is dominated by the S/N of the primary eclipse for the cases with low dilution (secondary S/N of 5 and 7). The curves with secondary S/N = 2 show the opposite trend --i.e. data support increasing for larger $q$-- because the primary eclipse S/N varies proportionally less in this case.

The light curves of planetary transits give a varying level of support for the PLANET model over the BEB model, depending on the radius of the planet, the impact parameter and the transit S/N (Fig.~\ref{fig.resultsPLANET}). The Bayes factor conclusively support the PLANET model if the transit is (close to) central and the transit signal-to-noise ratio is higher than about 50 - 100. For a given S/N and impact parameter, the Bayes factor of smaller planets is larger because the short ingress/engress times of the transit are difficult to reproduce by the BEB model. A systematic effect in the light curve, located close to the transit ingress for $b = 0.5$ provoke a strong decrease in the support for the PLANET hypothesis for $b = 0.5$, and hinders the interpretation of the dependence of the Bayes factor with impact parameter. For $b = 0.75$, only Earth-sized or Neptune-sized planets with high-S/N transits are supported strongly by the data.

For Earth-size planets, we computed further the Bayes factor between the PLANET model and models representing other false positive hypothesis. We found that triple hierarchical systems are discarded by the data, as already noted, for example, by \citet{torres2011}. On the other hand, the scenario consisting of a background star hosting a transiting (giant) planet whose light is diluted by the target star, received equal, or slightly stronger, support than the correct PLANET model. Similarly, the Bayes factor between the PLANET model and the model including a transiting planet orbiting the secondary component of a wide-orbit binary is too close to unity to allow preference for one model over the other. This is true even for transit light curves with S/N = 150.

Part b) of the analysis shows that the hypotheses prior odds, computed for conditions typical of \kepler\ candidates, strongly favor the PLANET model over the BEB model. This is mainly due to the strong constrain brought forth by the adaptive optics follow-up observations. As a consequence, the final odds ratio for the PLANET hypothesis based on planetary light curves is above 150 for all the planet scenarios (Fig.~\ref{fig.oddsratioPLA}), with the exception of the scenarios with $b=0.5$, affected by the systematic effect discussed above, for which the S/N needs to be above 100 - 150. On the contrary, the odds ratio in favour of the BEB model based on BEB light curves is consequently reduced. BEB scenarios with low and intermediate dilution level can still be correctly identified, but for scenarios with secondary S/N = 2 the odds ratio  does not conclusively support one model over the other (Fig.~\ref{fig.oddsratioBEB}).

Given this result, one might wonder if it is possible for an actual BEB to be identified as a transiting planet because the prior odds $\frac{\prob{PLANET}{I}}{\prob{BEB}{I}}$ is large enough. For example, if field around the target was less crowded that the one assumed here, or if the AO contrast curve was stronger, this might occur. However, one might argue that actual BEBs could not produce in this case the follow-up observations (in particular the AO contrast curve) used to compute the prior odds. In any case, our simulations coupled with the computation of the hypotheses ratios show the relative weight given by the Bayesian model comparison to data and priors.

Furthermore, we have shown that radial velocity follow-up observations can lead to the correct identification of the cases where the light curve alone does not suffice to conclude, even if a measurement of the mass of the transiting object is unattainable. Indeed, many of these false positive scenarios exhibit radial-velocity and bisector signals that could be detected with a relatively small number of RV observations, compared to those that would be needed to detect the reflex motion of the star \citep[e.g.][]{queloz2009}. Because the RV and bisector signal produced by a false positive can occur at any moment of the orbit, depending on the relative velocities of the target star and the blended system, the scheduling constraints usually associated with the follow-up of transiting candidates become less stringent. The background transiting planet scenario exhibits RV signals less frequently than other scenarios. This kind of false positive will certainly prove among the hardest to discard. In any case, our results underline the importance of intensive follow-up observations of transiting candidates, in particular of ground-based velocimetry measurements.

PASTIS has already been employed for analyzing datasets of transiting planets and brown dwarfs \citep[e.g.][]{hebrard2013, diaz2013}, and to validate transiting candidates for which no reflex motion of the parent star is detected (such as CoRoT-22 b, Moutou et al. submitted). It is currently being used to study real unresolved candidates from the \emph{CoRoT} space mission. A thorough comparison with BLENDER and the \citetalias{morton2012} procedure, based on the analysis of already validated candidates, will be presented in a forthcoming paper.

We have already identified a few features of PASTIS that will be improved in the future.
The treatment of systematic errors in the data might be an important issue to deal with in order to improve the method presented here. For the moment, systematic effects are treated as a source of additional gaussian noise whose amplitude is a model parameter. However, the example of the systematic feature affecting planetary light curves with $b = 0.5$ shows that this simple model is not sufficient. It also highlights the interest of performing these simulations using real data. A more realistic error distribution modeling, using techniques such as the autoregressive-moving-average model \citep[e.g.][]{tuomi2013} could permit detecting more subtle effects in the available data, and lead to a more robust determination of the Bayesian odds ratio. Another example of the need of a more sophisticated noise model is the case of Kepler-68 c \citep{gilliland2013}, whose validation is conditional to the nature of a small eclipse exactly in opposite phase to the transits. The authors claim that similar features are present in the light curve, which would render that particular "eclipse" non-significant. An adequate noise model could permit to quantify this statement.

Some observations that are usually available for transiting candidates, such as high-angular resolution imaging, or centroid motion --as provided, for example, by the \kepler\ pipeline-- are currently not modeled by PASTIS as is done for the light curves, radial velocities, etc. For the time being, PASTIS includes the information provided by these datasets in the prior odds computation, but self-consistently modeling these data is envisaged. This should increase the robustness of our determination of the evidences used for model comparison.

Future space missions such as PLATO will provide transits of small-size planet candidates at very high signal-to-noise ratio, due to the brightness of their target stars. Fully exploiting these data for statistical validation will require detailed physical modeling of the light curve. We have shown that PASTIS should be able to validate these candidates. Subsequent ground-based radial-velocity observations, focused on already validated candidates, would provide a measurement of their mass. Combined with the precise measurement of the radius from the space-based discovery light curve, the bulk density of Earth-size objects would be known with unprecedented precision.

\section*{acknowledgements}

The authors thankfully acknowledge the following individuals: Christian Marinoni for his valuable advise on Bayesian statistics and model comparison, Carlos L\'azaro for interesting discussions on eclipsing binaries, and Sergey Rodionov for his help in testing and improving the MCMC algorithm presented in this article. We are thankful to the LAM scientific computing team (CeSAM), in particular Thomas Fenouillet, for its assistance in using the computing cluster. We are also thankful to R. Sanchis-Ojeda for a fruitful discussion about the $b = 0.5$ effect, and the referee, Tim Morton, for a thorough and insightful report that led to an improved manuscript. RFD was supported by CNES via its postdoctoral fellowship program. JMA is supported by CNES. AS acknowledges the support by the European Research Council/European Community under the FP7 through Starting Grant agreement number 239953.

\input{PASTIS.bbl}
%\bibliographystyle{mn2e_corrected} 
%\bibliography{biblio_iafe}

\appendix

\section{Tables}
\begin{table*}
\center 
\caption{Results based on the the transiting planet synthetic data. \label{table.resultsPLA}}
\begin{tabular}{ccc|ll|cccc|ccc|cc}
\hline
\hline
\multicolumn{3}{c|}{Model} & \multicolumn{2}{c|}{\# Indep.\ samples} & \multicolumn{4}{c|}{Bayes factor / Odds ratio [$\log_{10}$]} & $\chi^2_\mathrm{PLANET}$ &$\chi^2_\mathrm{BEB}$&F-test & $D$ & $p$-value\\
b & snr & Rpl  & PLANET & BEB & $B_\mathrm{PB}$ & $OR_\mathrm{PB}$ & $\sigma_+$ & $\sigma_-$ &&&prob.&\\
 & & [R$_\oplus$] &&&&&[dex]&[dex]\\
\hline 
\input{tablePLANET.txt}

\hline
\hline
\end{tabular}
\end{table*}

%%%%%%%%%%
%% TABLES BEB
%%%%%%%%%%
\begin{table*}
\center 
\caption{Results based on the BEB synthetic data. \label{table.resultsBEB}}
\begin{tabular}{cccc|ll|cccc|ccc|cc}
\hline
\hline
\multicolumn{4}{c|}{Model} & \multicolumn{2}{c|}{\# Indep.\ samples} & \multicolumn{4}{c|}{Bayes factor / Odds ratio [$\log_{10}$]} & $\chi^2_\mathrm{BEB}$ &$\chi^2_\mathrm{PLANET}$&F-test & $D$ & $p$-value\\
b & S/N$_\mathrm{sec}$ & S/N$_\mathrm{pri}$ & q  & BEB & PLANET & $B_\mathrm{BP}$ & $OR_\mathrm{BP}$ & $\sigma_+$ & $\sigma_-$ &&&prob.&\\
 & & &&&&&&[dex]&[dex]\\
\hline 
\input{tableBEB.txt}
\hline
\hline
\end{tabular}
\end{table*}

\end{document}

%% file: tablePLANET.txt
0.00 &	 10 &	 1.0 &	 100989 &	 4858 &	0.45 &	4.01 &	0.02 &	0.06 &	0.93201 &	0.93480 &	0.44 &	1.40 &	0.92\\ 
0.00 &	 10 &	 4.4 &	 120491 &	 6598 &	0.39 &	2.79 &	0.02 &	0.06 &	0.93554 &	0.93324 &	0.55 &	0.68 &	0.98\\ 
0.00 &	 10 &	 7.8 &	 136852 &	 8342 &	0.20 &	2.95 &	0.02 &	0.13 &	0.93106 &	0.93212 &	0.48 &	2.37 &	0.80\\ 
0.00 &	 10 &	 11.2 &	 138711 &	 6493 &	0.07 &	3.30 &	0.04 &	0.06 &	0.93283 &	0.93602 &	0.43 &	1.49 &	0.91\\ 
0.00 &	 20 &	 1.0 &	 85846 &	 2112 &	0.93 &	4.49 &	0.09 &	0.10 &	0.93679 &	0.92976 &	0.65 &	2.10 &	0.84\\ 
0.00 &	 20 &	 4.4 &	 114168 &	 2360 &	0.71 &	3.11 &	0.05 &	0.07 &	0.93525 &	0.93631 &	0.48 &	1.95 &	0.85\\ 
0.00 &	 20 &	 7.8 &	 107104 &	 2319 &	0.63 &	3.38 &	0.03 &	0.07 &	0.93388 &	0.93213 &	0.54 &	0.68 &	0.98\\ 
0.00 &	 20 &	 11.2 &	 104127 &	 3254 &	0.57 &	3.79 &	0.05 &	0.10 &	0.93618 &	0.93125 &	0.60 &	0.54 &	0.99\\ 
0.00 &	 50 &	 1.0 &	 54916 &	 1450 &	1.20 &	4.75 &	0.04 &	0.06 &	0.93397 &	0.93337 &	0.51 &	-1.99 &	---\\ 
0.00 &	 50 &	 4.4 &	 71055 &	 1398 &	1.21 &	3.61 &	0.05 &	0.07 &	0.93373 &	0.93087 &	0.56 &	-3.48 &	---\\ 
0.00 &	 50 &	 7.8 &	 89813 &	 1510 &	1.68 &	4.43 &	0.05 &	0.08 &	0.93875 &	0.93623 &	0.55 &	-4.57 &	---\\ 
0.00 &	 50 &	 11.2 &	 104625 &	 2469 &	1.73 &	4.95 &	0.04 &	0.07 &	0.93840 &	0.93558 &	0.56 &	-6.24 &	---\\ 
0.00 &	 100 &	 1.0 &	 4602 &	 1558 &	7.80 &	11.36 &	0.05 &	0.09 &	0.93311 &	0.94038 &	0.35 &	-31.52 &	---\\ 
0.00 &	 100 &	 4.4 &	 29266 &	 1227 &	5.98 &	8.38 &	0.10 &	0.12 &	0.93372 &	0.93051 &	0.57 &	-24.75 &	---\\ 
0.00 &	 100 &	 7.8 &	 52806 &	 1385 &	5.05 &	7.79 &	0.04 &	0.16 &	0.93832 &	0.93668 &	0.54 &	-20.54 &	---\\ 
0.00 &	 100 &	 11.2 &	 77357 &	 1495 &	6.40 &	9.62 &	0.06 &	0.08 &	0.93700 &	0.94299 &	0.38 &	-27.25 &	---\\ 
0.00 &	 150 &	 1.0 &	 2607 &	 4186 &	19.59 &	23.15 &	0.07 &	0.09 &	0.93367 &	0.94215 &	0.33 &	-85.46 &	---\\ 
0.00 &	 150 &	 4.4 &	 23076 &	 5406 &	13.78 &	16.18 &	0.07 &	0.10 &	0.93847 &	0.94395 &	0.39 &	-59.42 &	---\\ 
0.00 &	 150 &	 7.8 &	 49800 &	 1030 &	10.54 &	13.29 &	0.07 &	0.11 &	0.93889 &	0.93692 &	0.54 &	-44.84 &	---\\ 
0.00 &	 150 &	 11.2 &	 72934 &	 2103 &	13.23 &	16.46 &	0.08 &	0.12 &	0.93685 &	0.92556 &	0.73 &	-58.03 &	---\\ 
\hline
0.50 &	 10 &	 1.0 &	 112310 &	 6268 &	-0.35 &	3.20 &	0.03 &	0.05 &	0.93910 &	0.94034 &	0.47 &	1.04 &	0.96\\ 
0.50 &	 10 &	 4.4 &	 124581 &	 9092 &	-0.37 &	2.03 &	0.01 &	0.08 &	0.93447 &	0.93599 &	0.47 &	2.34 &	0.80\\ 
0.50 &	 10 &	 7.8 &	 139455 &	 8309 &	-0.40 &	2.35 &	0.02 &	0.08 &	0.93494 &	0.93583 &	0.48 &	2.72 &	0.74\\ 
0.50 &	 10 &	 11.2 &	 136480 &	 8852 &	-0.43 &	2.79 &	0.02 &	0.05 &	0.93667 &	0.92381 &	0.76 &	3.14 &	0.68\\ 
0.50 &	 20 &	 1.0 &	 104871 &	 3289 &	-1.14 &	2.42 &	0.04 &	0.06 &	0.93540 &	0.93245 &	0.56 &	7.50 &	0.19\\ 
0.50 &	 20 &	 4.4 &	 108591 &	 2394 &	-1.10 &	1.30 &	0.03 &	0.05 &	0.93758 &	0.94582 &	0.33 &	6.52 &	0.26\\ 
0.50 &	 20 &	 7.8 &	 114044 &	 3790 &	-0.81 &	1.94 &	0.05 &	0.06 &	0.93636 &	0.93925 &	0.44 &	6.60 &	0.25\\ 
0.50 &	 20 &	 11.2 &	 107599 &	 2782 &	-0.71 &	2.51 &	0.05 &	0.06 &	0.93458 &	0.92945 &	0.61 &	4.27 &	0.51\\ 
0.50 &	 50 &	 1.0 &	 32748 &	 1100 &	-1.13 &	2.43 &	0.04 &	0.10 &	0.93613 &	0.93090 &	0.61 &	3.91 &	0.56\\ 
0.50 &	 50 &	 4.4 &	 20593 &	 1037 &	-1.68 &	0.72 &	0.04 &	0.07 &	0.93505 &	0.92921 &	0.62 &	7.54 &	0.18\\ 
0.50 &	 50 &	 7.8 &	 43817 &	 1523 &	-1.90 &	0.85 &	0.03 &	0.08 &	0.94088 &	0.94298 &	0.46 &	6.42 &	0.27\\ 
0.50 &	 50 &	 11.2 &	 51981 &	 2022 &	-1.58 &	1.65 &	0.04 &	0.08 &	0.93109 &	0.93701 &	0.38 &	5.06 &	0.41\\ 
0.50 &	 100 &	 1.0 &	 9408 &	 1776 &	0.47 &	4.03 &	0.04 &	0.06 &	0.94104 &	0.93015 &	0.72 &	0.45 &	0.99\\ 
0.50 &	 100 &	 4.4 &	 15604 &	 3074 &	-0.46 &	1.94 &	0.05 &	0.08 &	0.93651 &	0.93251 &	0.58 &	3.26 &	0.66\\ 
0.50 &	 100 &	 7.8 &	 24370 &	 2025 &	-1.13 &	1.61 &	0.04 &	0.08 &	0.92973 &	0.94368 &	0.23 &	4.52 &	0.48\\ 
0.50 &	 100 &	 11.2 &	 60682 &	 1328 &	-1.17 &	2.06 &	0.05 &	0.08 &	0.93869 &	0.93768 &	0.52 &	4.12 &	0.53\\ 
0.50 &	 150 &	 1.0 &	 3934 &	 1663 &	5.76 &	9.32 &	0.09 &	0.09 &	0.93617 &	0.93756 &	0.47 &	-24.11 &	---\\ 
0.50 &	 150 &	 4.4 &	 10341 &	 1385 &	2.18 &	4.58 &	0.09 &	0.10 &	0.93406 &	0.94182 &	0.34 &	-7.88 &	---\\ 
0.50 &	 150 &	 7.8 &	 31760 &	 1705 &	-0.06 &	2.68 &	0.04 &	0.06 &	0.93941 &	0.93767 &	0.54 &	1.26 &	0.94\\ 
0.50 &	 150 &	 11.2 &	 65257 &	 1173 &	0.35 &	3.57 &	0.07 &	0.09 &	0.93728 &	0.94398 &	0.36 &	-1.17 &	---\\ 
\hline
0.75 &	 10 &	 1.0 &	 72652 &	 8919 &	-0.58 &	2.98 &	0.03 &	0.06 &	0.93686 &	0.93962 &	0.44 &	1.66 &	0.89\\ 
0.75 &	 10 &	 4.4 &	 79115 &	 5611 &	-0.65 &	1.75 &	0.04 &	0.06 &	0.94027 &	0.93326 &	0.65 &	2.12 &	0.83\\ 
0.75 &	 10 &	 7.8 &	 84596 &	 7868 &	-0.73 &	2.01 &	0.04 &	0.06 &	0.93593 &	0.93926 &	0.43 &	4.16 &	0.53\\ 
0.75 &	 10 &	 11.2 &	 66029 &	 9165 &	-0.82 &	2.40 &	0.04 &	0.07 &	0.93489 &	0.93765 &	0.44 &	4.74 &	0.45\\ 
0.75 &	 20 &	 1.0 &	 121978 &	 3093 &	0.84 &	4.40 &	0.04 &	0.07 &	0.93679 &	0.93699 &	0.50 &	2.69 &	0.75\\ 
0.75 &	 20 &	 4.4 &	 130159 &	 3521 &	0.62 &	3.02 &	0.04 &	0.06 &	0.93390 &	0.93987 &	0.38 &	2.05 &	0.84\\ 
0.75 &	 20 &	 7.8 &	 110822 &	 4129 &	0.35 &	3.09 &	0.03 &	0.08 &	0.93557 &	0.93523 &	0.51 &	0.27 &	1.00\\ 
0.75 &	 20 &	 11.2 &	 124580 &	 4031 &	-0.08 &	3.15 &	0.02 &	0.09 &	0.93623 &	0.92781 &	0.67 &	1.43 &	0.92\\ 
0.75 &	 50 &	 1.0 &	 110823 &	 1051 &	1.64 &	5.20 &	0.07 &	0.10 &	0.93344 &	0.93572 &	0.45 &	-4.31 &	---\\ 
0.75 &	 50 &	 4.4 &	 117143 &	 2695 &	0.71 &	3.11 &	0.06 &	0.10 &	0.93405 &	0.94063 &	0.36 &	0.59 &	0.99\\ 
0.75 &	 50 &	 7.8 &	 121605 &	 1539 &	0.12 &	2.86 &	0.05 &	0.07 &	0.93334 &	0.93226 &	0.52 &	1.92 &	0.86\\ 
0.75 &	 50 &	 11.2 &	 75865 &	 3027 &	0.74 &	3.96 &	0.04 &	0.10 &	0.93796 &	0.93438 &	0.58 &	0.59 &	0.99\\ 
0.75 &	 100 &	 1.0 &	 91985 &	 6659 &	6.95 &	10.51 &	0.05 &	0.11 &	0.93651 &	0.93583 &	0.51 &	-26.80 &	---\\ 
0.75 &	 100 &	 4.4 &	 105617 &	 3390 &	2.42 &	4.82 &	0.06 &	0.14 &	0.93937 &	0.92997 &	0.69 &	-6.38 &	---\\ 
0.75 &	 100 &	 7.8 &	 96691 &	 3436 &	0.33 &	3.08 &	0.02 &	0.13 &	0.93566 &	0.92618 &	0.69 &	0.43 &	0.99\\ 
0.75 &	 100 &	 11.2 &	 107104 &	 3073 &	0.64 &	3.86 &	0.04 &	0.07 &	0.94024 &	0.94558 &	0.39 &	-1.64 &	---\\ 
0.75 &	 150 &	 1.0 &	 85042 &	 9325 &	15.71 &	19.27 &	0.06 &	0.15 &	0.93476 &	0.93627 &	0.47 &	-66.96 &	---\\ 
0.75 &	 150 &	 4.4 &	 107104 &	 5299 &	5.72 &	8.12 &	0.09 &	0.17 &	0.93810 &	0.94018 &	0.46 &	-21.56 &	---\\ 
0.75 &	 150 &	 7.8 &	 106138 &	 1737 &	1.10 &	3.85 &	0.06 &	0.09 &	0.93827 &	0.93672 &	0.53 &	-2.22 &	---\\ 
0.75 &	 150 &	 11.2 &	 95204 &	 1051 &	1.64 &	4.86 &	0.04 &	0.08 &	0.94156 &	0.94707 &	0.39 &	-5.68 &	---\\

%% file: tableBEB.txt
0.00 &	 2.0 &	 106 &	 0.1 &	 1771 &	 105616 &	1.81 &	-1.38 &	0.06 &	0.16 &	0.92851 &	0.93720 &	0.32 &	16.26 &	6.2e-03\\ 
0.00 &	 2.0 &	 89 &	 0.3 &	 3946 &	 100285 &	2.39 &	-1.10 &	0.05 &	0.12 &	0.93804 &	0.93412 &	0.58 &	11.78 &	3.8e-02\\ 
0.00 &	 2.0 &	 65 &	 0.5 &	 8211 &	 106360 &	2.32 &	-1.16 &	0.05 &	0.14 &	0.93563 &	0.93505 &	0.51 &	9.94 &	7.7e-02\\ 
0.00 &	 5.0 &	 266 &	 0.1 &	 3158 &	 31920 &	13.00 &	9.81 &	0.09 &	0.29 &	0.93787 &	0.93068 &	0.65 &	66.93 &	4.5e-13\\ 
0.00 &	 5.0 &	 223 &	 0.3 &	 3010 &	 17043 &	6.04 &	2.84 &	0.10 &	0.23 &	0.93847 &	0.93773 &	0.52 &	33.48 &	3.0e-06\\ 
0.00 &	 5.0 &	 162 &	 0.5 &	 2439 &	 24697 &	6.07 &	2.88 &	0.14 &	0.19 &	0.93231 &	0.93709 &	0.40 &	26.38 &	7.5e-05\\ 
0.00 &	 7.0 &	 373 &	 0.1 &	 2651 &	 1884 &	21.83 &	18.64 &	0.08 &	0.23 &	0.94135 &	0.94273 &	0.47 &	93.81 &	0.0e+00\\ 
0.00 &	 7.0 &	 313 &	 0.3 &	 2637 &	 98230 &	11.05 &	7.85 &	0.09 &	0.20 &	0.93195 &	0.93298 &	0.48 &	54.08 &	2.0e-10\\ 
0.00 &	 7.0 &	 227 &	 0.5 &	 1511 &	 14681 &	9.00 &	5.81 &	0.09 &	0.16 &	0.94476 &	0.93913 &	0.62 &	42.10 &	5.6e-08\\ 
\hline
0.50 &	 2.0 &	 100 &	 0.1 &	 1370 &	 104625 &	2.38 &	-0.81 &	0.05 &	0.15 &	0.92869 &	0.93067 &	0.46 &	14.94 &	1.1e-02\\ 
0.50 &	 2.0 &	 83 &	 0.3 &	 4621 &	 44229 &	3.21 &	-0.27 &	0.04 &	0.16 &	0.93544 &	0.93497 &	0.51 &	14.52 &	1.3e-02\\ 
0.50 &	 2.0 &	 59 &	 0.5 &	 9967 &	 73582 &	2.50 &	-0.98 &	0.04 &	0.13 &	0.93787 &	0.93228 &	0.62 &	10.54 &	6.1e-02\\ 
0.50 &	 5.0 &	 251 &	 0.1 &	 2284 &	 91984 &	11.69 &	8.50 &	0.08 &	0.21 &	0.94182 &	0.93866 &	0.57 &	58.94 &	2.0e-11\\ 
0.50 &	 5.0 &	 208 &	 0.3 &	 1828 &	 1661 &	6.40 &	3.21 &	0.18 &	0.22 &	0.93956 &	0.92872 &	0.72 &	26.90 &	6.0e-05\\ 
0.50 &	 5.0 &	 148 &	 0.5 &	 4322 &	 2944 &	5.82 &	2.63 &	0.15 &	0.20 &	0.93812 &	0.93850 &	0.49 &	23.14 &	3.2e-04\\ 
0.50 &	 7.0 &	 352 &	 0.1 &	 2128 &	 20641 &	20.74 &	17.55 &	0.08 &	0.24 &	0.94275 &	0.93920 &	0.57 &	91.31 &	0.0e+00\\ 
0.50 &	 7.0 &	 292 &	 0.3 &	 1889 &	 13937 &	9.15 &	5.96 &	0.03 &	0.33 &	0.93087 &	0.93965 &	0.32 &	45.07 &	1.4e-08\\ 
0.50 &	 7.0 &	 208 &	 0.5 &	 3500 &	 43210 &	8.41 &	5.22 &	0.08 &	0.15 &	0.92630 &	0.93792 &	0.27 &	37.97 &	3.8e-07\\ 
\hline
0.75 &	 2.0 &	 88 &	 0.1 &	 1874 &	 119375 &	1.83 &	-1.36 &	0.05 &	0.14 &	0.94558 &	0.93112 &	0.78 &	16.82 &	4.9e-03\\ 
0.75 &	 2.0 &	 71 &	 0.3 &	 7782 &	 99284 &	2.25 &	-1.23 &	0.06 &	0.13 &	0.93433 &	0.93419 &	0.50 &	10.96 &	5.2e-02\\ 
0.75 &	 2.0 &	 53 &	 0.5 &	 7371 &	 85539 &	2.38 &	-1.11 &	0.04 &	0.14 &	0.93973 &	0.93857 &	0.53 &	12.63 &	2.7e-02\\ 
0.75 &	 5.0 &	 222 &	 0.1 &	 1937 &	 62499 &	11.55 &	8.36 &	0.06 &	0.21 &	0.92450 &	0.93293 &	0.33 &	54.15 &	1.9e-10\\ 
0.75 &	 5.0 &	 177 &	 0.3 &	 2366 &	 3849 &	5.19 &	2.00 &	0.11 &	0.16 &	0.93712 &	0.93276 &	0.59 &	23.15 &	3.2e-04\\ 
0.75 &	 5.0 &	 132 &	 0.5 &	 8816 &	 98924 &	7.19 &	3.99 &	0.04 &	0.16 &	0.93341 &	0.93481 &	0.47 &	33.68 &	2.8e-06\\ 
0.75 &	 7.0 &	 311 &	 0.1 &	 1525 &	 37110 &	19.13 &	15.93 &	0.08 &	0.35 &	0.92688 &	0.93736 &	0.29 &	85.93 &	0.0e+00\\ 
0.75 &	 7.0 &	 249 &	 0.3 &	 4397 &	 68283 &	7.85 &	4.65 &	0.08 &	0.18 &	0.93759 &	0.93526 &	0.55 &	38.16 &	3.5e-07\\ 
0.75 &	 7.0 &	 185 &	 0.5 &	 7654 &	 11715 &	7.93 &	4.74 &	0.06 &	0.17 &	0.93521 &	0.93271 &	0.55 &	34.68 &	1.7e-06\\ 